\documentclass[final,5p,times,twocolumn,sort&compress]{elsarticle}

\usepackage{amssymb}
\usepackage{amsfonts}
\usepackage{amsmath}
\usepackage[american]{babel}
\usepackage{hyperref}
\usepackage[utf8]{inputenc}
\usepackage[T1]{fontenc}
\usepackage{lmodern}

\newcommand{\norm}{\mathcal{P}}
\begin{hyphenrules}{american}
\end{hyphenrules}

\DeclareMathOperator{\sech}{sech}

\journal{Physica D}

\begin{document}

\begin{frontmatter}


\title{Modulational instability and localized breather modes 
\\in the discrete nonlinear Schrödinger equation with helicoidal hopping}


\author[zoq]{J. Stockhofe\corref{cor1}}
\cortext[cor1]{Corresponding author.}
\ead{jstockho@physnet.uni-hamburg.de}
\address[zoq]{Zentrum f\"ur Optische Quantentechnologien,
  Universit\"at Hamburg, Luruper Chaussee 149, 22761 Hamburg, Germany}

\author[zoq,cui]{P. Schmelcher}
\ead{pschmelc@physnet.uni-hamburg.de}
\address[cui]{The Hamburg Centre for Ultrafast Imaging, Universit\"at Hamburg, Luruper Chaussee 149, 22761 Hamburg, Germany}

\begin{abstract}
We study a one-dimensional discrete nonlinear Schrödinger model with hopping to the first and a selected $N$-th neighbor,
motivated by a helicoidal arrangement of lattice sites.
We provide a detailed analysis of the modulational instability properties of this equation,
identifying distinctive multi-stage instability cascades due to the helicoidal hopping term.
Bistability is a characteristic feature of the intrinsically localized breather modes,
and it is shown that information on the stability properties of weakly localized solutions 
can be inferred from the plane-wave modulational instability results.
Based on this argument, we derive analytical estimates of the critical parameters at which
the fundamental on-site breather branch of solutions turns unstable.
In the limit of large $N$, these estimates predict the emergence of an effective threshold behavior,
which can be viewed as the result of a dimensional crossover to a two-dimensional square lattice.
\end{abstract}

\end{frontmatter}


\section{Introduction}
The discrete nonlinear Schrödinger (DNLS) equation is one of the most widely studied models of nonlinear dynamics,
finding immediate modern applications in the description of optical waveguide arrays or ultracold bosonic atoms in periodic potentials,
but also arising as the fundamental envelope approximation to a vast diversity of lattice models 
from various fields of physics \cite{Kevrekidis2009}.
In its one-dimensional (1D) version, the DNLS equation models the dispersive excitation transfer on a discrete chain
in the presence of on-site nonlinearity.
Traditionally, the dispersive coupling, or ``hopping'', is restricted to nearest neighbors (NN) here,
which is typically justified by the increasing spatial distance to more remote neighbors 
causing a suppression of the higher-order hopping.
Conversely, the strong sensitivity of the coupling terms to the inter-site distances
can be employed to tune them via deformations of the lattice geometry, e.g. by introducing a bend into a 1D chain of sites
\cite{Gaididei2000,Christiansen2001,Archilla2001,Archilla2002,Cuevas2002,Kivshar2003}.
If the couplings originate from anisotropic (such as dipolar) interaction, the long-range terms can also be tuned
via the orientations of the local oscillators \cite{Sanchez-Rey2002}.
A specific proposal for enhancing the second-neighbor hopping by arranging the lattice sites in a zigzag structure \cite{Efremidis2002}
stimulated numerous recent theoretical and experimental investigations in this direction,
covering free expansion \cite{Dreisow2008} and Bloch oscillation \cite{Wang2010,Dreisow2011,Stockhofe2015} dynamics,
disorder-induced localization \cite{Golshani2013}, defect scattering \cite{Stockhofe2015a},
nonlinear localized excitations \cite{Efremidis2002,Kevrekidis2003,Szameit2009,Kevrekidis2009a,Chong2011} and self-trapping \cite{Yan2013} in such lattices.
Generalizing this idea, a three-dimensional layout of lattice sites along a helix curve
will lead to small inter-site distances (and thus large hopping probabilities) not only to the neighboring sites along the curve,
but also to certain sites on the adjacent windings of the helix,
admitting a geometry-induced enhancement of selected $N$-th neighbor hopping terms \cite{Stockhofe2015}, see also \cite{Wang1991,Xiong1992}.
The recently proposed experimental implementation of helix-shaped optical traps for ultracold atoms \cite{Reitz2012} calls for
a deeper understanding of the bosonic quantum many-body physics in helical geometries.
Within mean-field theory, this is formulated as a nonlinear Schrödinger problem \cite{Pethick2008}
such that a DNLS equation extended by an $N$-th neighbor inter-winding hopping term is a suitable model for condensed bosons in helix lattices.
While the focus is on neutral atoms here, it is worth noting that also particles with long-range (Coulomb or dipolar) interaction constrained to move on helix curves are subject to active research
\cite{Schmelcher2011,Zampetaki2013,Zampetaki2015,Zampetaki2015a,Law2008,Pedersen2014,Pedersen2016}.

Beyond this, the interplay of helical geometry, discreteness and nonlinearity has a long-standing history in the modeling of biomolecules, and in particular of the DNA double-helix molecule.
There are ongoing efforts to devise and refine tractable nonlinear lattice models that adequately describe the various functionally relevant aspects of the DNA dynamics \cite{Yakushevich2004}.
In this context, inter-winding coupling terms induced by the helix geometry have been argued to be significant, 
most prominently in the extension of the so-called Peyrard-Bishop (PB) model \cite{Peyrard1989} for DNA opening put forward in \cite{Gaeta1990,Dauxois1991,Dauxois1991a}.
The resulting ``helicoidal'' PB model has been extensively analyzed \cite{Gaeta1993,Zdravkovic2007,Zdravkovic2007a,Tabi2008,Gaeta2008,Tabi2009b,DangKoko2012}, 
see \cite{Zdravkovic2011} for a review.
Remarkably, under certain simplifying assumptions, the PB model reduces to a 1D-DNLS system,
and the helicoidal inter-winding coupling term turns into an isolated $N$-th neighbor hopping \cite{Mingaleev1999,Tabi2008},
providing an independent motivation for studying such extensions of the DNLS equation.

A key feature of many nonlinear lattice models is the possibility of spatially localized excitations
that do not disperse into the lattice, see \cite{Flach2008,Lederer2008} for recent reviews. 
Depending on the context, these have been
termed discrete solitons, intrinsic localized modes or breathers.
The basic breather solutions to the NN-restricted 1D-DNLS equation are well understood \cite{Kevrekidis2009},
but adding beyond-NN hopping terms to the model turns out to severely affect their properties.
In particular, second-neighbor hopping has been observed to induce bistability in the fundamental branch of breather modes, 
i.e. the coexistence of multiple such breather solutions at the same norm, but different frequencies \cite{Efremidis2002,Szameit2009},
while at the same time giving rise to an effective threshold norm below which only relatively broad localized solutions exist, 
as also evidenced in the experiment \cite{Szameit2009}. Similar findings had been reported before for 1D-DNLS systems with
higher powers of the on-site nonlinearity \cite{Laedke1994,Malomed1996,Cuevas2009a} or with long-range hopping of exponential or algebraic decay \cite{Gaididei1997,Johansson1998,Flach1998},
including also a proposal for switching between the coexisting breather solutions using external fields.

In this work, we provide a study of breather solutions in the 1D-DNLS equation extended by a helicoidal hopping term to a selected $N$-th neighbor,
such that the zigzag-geometry model of \cite{Efremidis2002} is included as a special case for $N=2$.
Generically, the existence of localized breather solutions is intimately linked to the so-called modulational instability (MI) of plane waves \cite{Kivshar1992,Smerzi2002,Kevrekidis2009},
and the MI properties of the zigzag model have been successfully used to predict the existence of different types of localized breathers \cite{Hennig2001,Efremidis2002}.
We argue that the plane-wave MI analysis can provide insight not only into the existence, but also into the stability properties of localized solutions in the helicoidal DNLS model,
admitting analytical estimates of the parameter values at which breather instability sets in.
This will be shown first for $N=2$ and then generalized to arbitrary $N$, demonstrating the versatile applicability of this type of argument.
Furthermore, we discuss the limit of large $N$, illustrating that it can be viewed as an effective dimensional crossover to a DNLS equation on a square lattice.
In this spirit, the breather bistability in the zigzag model can be understood as the precursor of the well-known threshold behavior in the two-dimensional DNLS equation \cite{Flach1997,Weinstein1999}.
The modulational-instability-based estimates are remarkably effective in predicting this crossover, and even give a reasonable approximation to the norm of the Townes soliton.

Our presentation is structured as follows: In Sec.~\ref{sec:setup} we introduce and motivate the helicoidal DNLS model.
Sec.~\ref{sec:MI} gives a discussion of the linear dispersion relation and a detailed analysis of the MI features.
The main results on breather stability properties and their connection to the plane-wave MI analysis are presented in Sec.~\ref{sec:br},
distinguishing the cases of attractive and repulsive nonlinearity.
Finally, we discuss the dimensional crossover to large $N$ in Sec.~\ref{sec:2D}, before concluding in Sec.~\ref{sec:con}.
Details on the variational and continuum approximations employed in the manuscript are given in appendix~\ref{app}.

\section{Setup}
\label{sec:setup}
We study a generalized DNLS model of the form
\begin{align}
i \frac{\text d \Psi_j}{\text d \tau}  = &-t_1(\Psi_{j+1} + \Psi_{j-1} ) \nonumber\\
&-t_N(\Psi_{j+N} + \Psi_{j-N} ) + U |\Psi_j|^2 \Psi_j,
\label{eq:dnls}
\end{align}
where $\Psi_j=\Psi_j(\tau)$ is the complex wave amplitude at site $j \in \mathbb Z$ and $\tau$ denotes time.
The parameter $U$ fixes the nonlinearity (attractive for $U<0$, repulsive for $U>0$).
We account for hopping terms to the first and a selected, fixed $N$-th neighbor ($N \geq 2$) with positive amplitudes
$t_1$ and $t_N$, respectively.
Dimensionless units are employed throughout.
We will keep all three parameters $t_1$, $t_N$ and $U$ for clarity, although by rescaling time and the norm of the $\Psi_j$ one could
set $t_1=1$, $|U|=1$ without loss of generality.
Our numerical simulations of Eq.~(\ref{eq:dnls}) are performed on finite domains of $L$ sites with periodic boundary conditions
where typically $L=500$ or $L=1000$, much larger than the extension of the localized solutions we are mainly interested in.

\begin{figure}[ht!]
 \includegraphics[height=3.5cm]{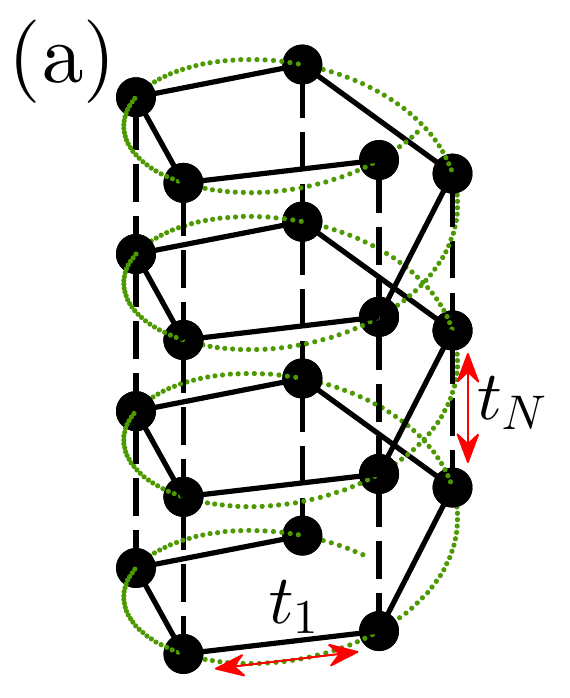} \includegraphics[height=3.5cm]{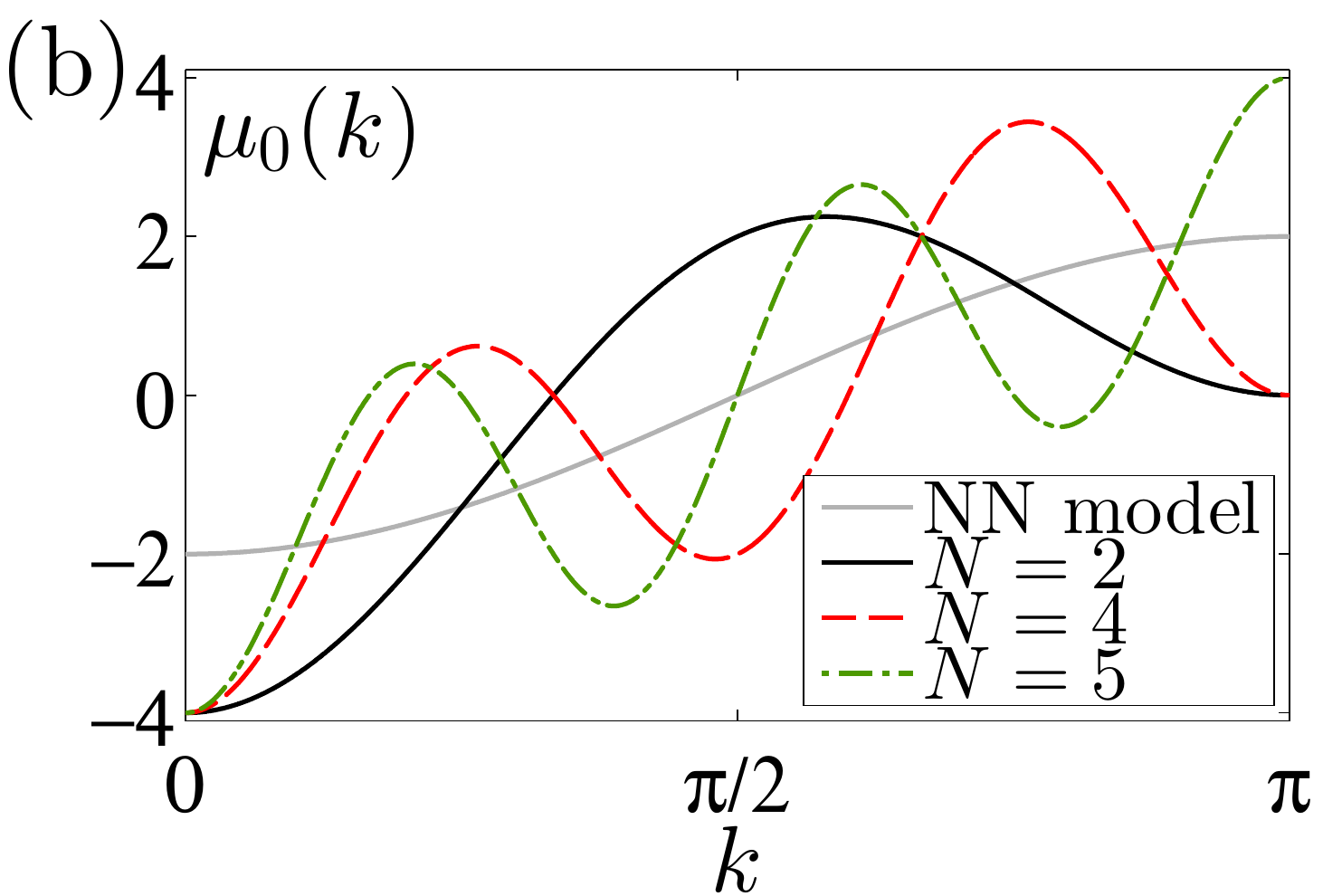}
\caption{\label{fig:setup}(Color online) (a) Sketch of a helicoidal lattice with $N=5$ sites per winding. The dotted green line depicts the underlying continuous helix curve. Full and dashed black lines indicate the nearest-neighbor hopping $t_1$ along the same winding
and the $N$-th neighbor hopping $t_N$ to the closest sites on the adjacent windings, respectively.  (b) Linear dispersion relations for the helicoidal DNLS equation with $t_1=1$, $t_N=0$ (NN model) and 
$t_1=t_N=1$ at different $N$.}\end{figure}
The unusual hopping structure of Eq.~(\ref{eq:dnls}) is thought to emerge in helicoidal arrangements of lattice sites
as sketched in Fig.~\ref{fig:setup}(a), with discrete sites placed equidistantly along a helix curve.
Counting the sites with integers $j$ along the curve, a given site will have particularly small distances (and thus a large hopping amplitude) to its nearest neighbors along the same winding (links indicated by solid lines),
but potentially also to certain sites on the adjacent windings, which are further apart in index, but close in three-dimensional space (links indicated by dashed lines).
For simplicity, we focus here on certain commensurate lattice geometries with $N$ sites per winding in which only hopping to a single site on each of the neighboring windings is accounted for.
Crucially, invariance of the helix lattice under discrete screw operations leads to an effective translational invariance of the model, in that the spatial difference between two sites only depends on their index difference \cite{Stockhofe2015},
justifying the use of site-independent hopping parameters $t_1$, $t_N$ in Eq.~(\ref{eq:dnls}).

Choosing $N=2$, our model coincides with the equation of motion put forward for light evolution in zigzag arrangements of evanescently coupled optical waveguides in \cite{Efremidis2002}, as implemented in \cite{Dreisow2008,Szameit2009,Dreisow2011}.
In this case, our discussion of the localized breather modes in Sec.~\ref{sec:br} directly relates to the experimental and numerical observations reported in \cite{Szameit2009}.
The extension to arbitrary values of $N$ is conceptually appealing as it puts the results for the zigzag lattice into a more abstract perspective, revealing also a 
connection to the well-understood two-dimensional DNLS model, see Sec.~\ref{sec:2D}. 
Experimentally, in the realm of optics the third spatial dimension enters the DNLS equation as the time variable, such that effectively only planar lattices can be designed. 
However, three-dimensional potential landscapes with deep minima at the desired helix lattice sites could be tailored for ultracold bosonic atoms, cf. \cite{Reitz2012,Stockhofe2015}
(see also \cite{Zhang2015} for a recent proposal of zigzag optical lattices).
In this framework, a DNLS description arises in the mean-field treatment of the lowest-band tight-binding model, 
with $\Psi_j$ denoting the local condensate order parameter at site $j$ \cite{Trombettoni2001,Cataliotti2001}.
Here, the effective nonlinearity originates from the interatomic contact interaction at low temperatures \cite{Pethick2008}.
The corresponding DNLS-type dynamics of ultracold bosons in one-dimensional optical lattices has been experimentally observed \cite{Cataliotti2001,Cataliotti2003}.
Furthermore, a variety of other nonlinear lattice models reduce to DNLS systems within suitable envelope approximation schemes \cite{Mingaleev1999},
which significantly enhances the range of applicability of DNLS-based results.
For instance, as noted above a DNLS equation with extended hopping as in Eq.~(\ref{eq:dnls}) has recently been shown to arise
in approximate treatments of the helicoidal Peyrard-Bishop model of DNA mechanics \cite{Tabi2008}.

\section{Modulational instability}
\label{sec:MI}
In this section, we prepare the analysis of localized breather solutions 
by providing a comprehensive discussion of plane wave solutions to the helicoidal DNLS equation and their stability in the presence of nonlinearity.
First, the stationary counterpart of Eq.~(\ref{eq:dnls}) is obtained in the usual way by factorizing $\Psi_j(\tau) = \psi_j \exp(-i \mu \tau)$,
yielding
\begin{align}
 \mu \psi_j = &-t_1(\psi_{j+1} + \psi_{j-1} ) \nonumber\\
&-t_N(\psi_{j+N} + \psi_{j-N} ) + U |\psi_j|^2 \psi_j
\label{eq:sdnls}
\end{align}
for the time-independent complex amplitudes $\psi_j$.
Eq.~(\ref{eq:sdnls}) possesses plane wave solutions of the form $\psi_j = A \exp(i k j)$, 
with a constant amplitude $A \in \mathbb C$ and $k$ denoting the quasi-momentum which can be restricted to the first Brillouin zone $-\pi < k \leq \pi$ and assumes continuous values for infinitely extended lattices as considered here.
Inserting the plane wave ansatz into Eq.~(\ref{eq:sdnls}) gives the relation $\mu(k)=\mu_0(k) + U|A|^2$, where
\begin{equation}
 \mu_0(k)=-2 t_1 \cos(k) - 2 t_N \cos(kN)
\label{eq:ldisp}
\end{equation}
is the dispersion relation in the absence of nonlinearity.
Fig.~\ref{fig:setup}(b) shows examples of dispersion curves for various values of the model parameters. While the NN model features a monotonous increase of $\mu_0$ from the lower band edge at $k=0$ to the upper band edge at $k=\pi$,
the $N$-th neighbor hopping term induces additional modulations and the emergence of nontrivial extrema inside the first Brillouin zone,
as illustrated in Fig.~\ref{fig:setup}(b) for $t_N=t_1$, but similarly present for other ratios of the hoppings. For $N$ odd, the dispersion relation still features the symmetry $\mu_0(k)=-\mu_0(\pi-k)$,
meaning in particular that the upper band edge remains at $k=\pi$. In contrast, for $N$ even this symmetry is lost and generically, for not too small $t_N$, the upper band edge is shifted away from $k=\pi$.
Thus, in this case there are two inequivalent global maxima $\pm k_m$ of $\mu_0$, which is referred to as a split (upper) band edge.

\begin{figure}[ht]
\includegraphics[width=0.45\textwidth]{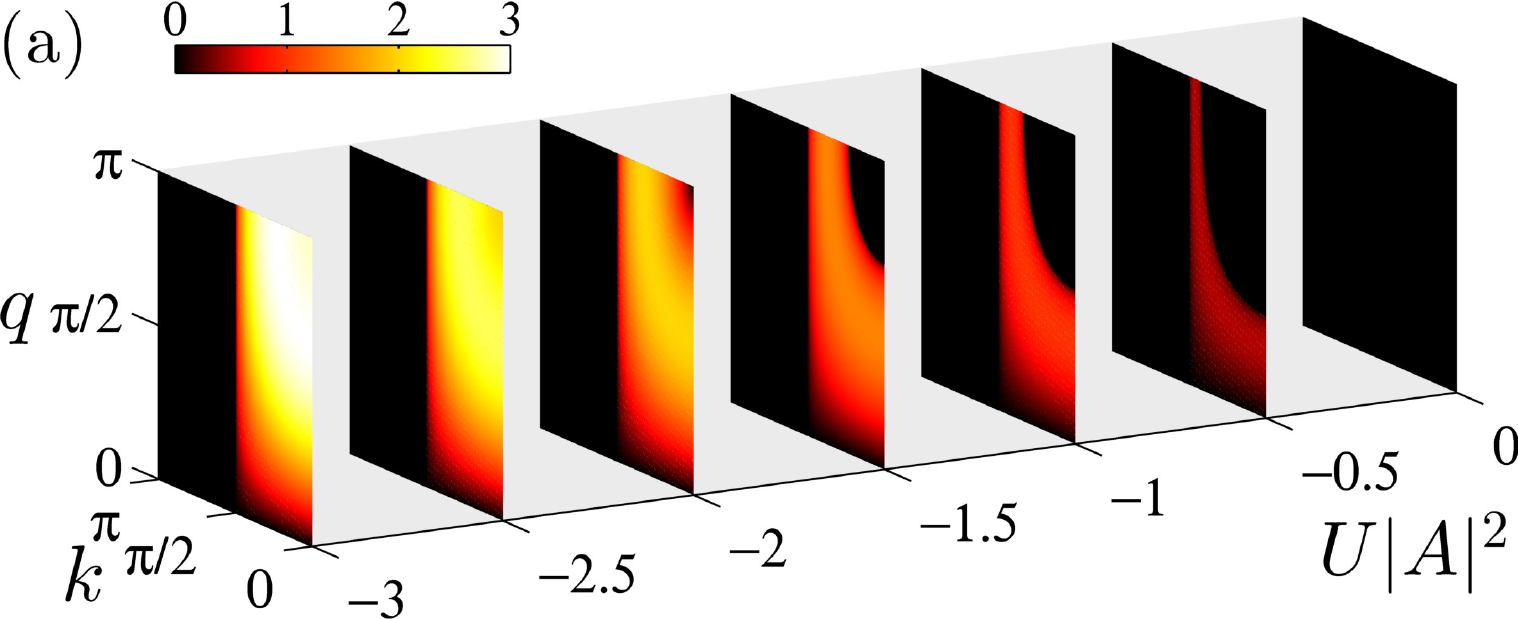}\\
  \vspace{2mm}\includegraphics[width=0.45\textwidth]{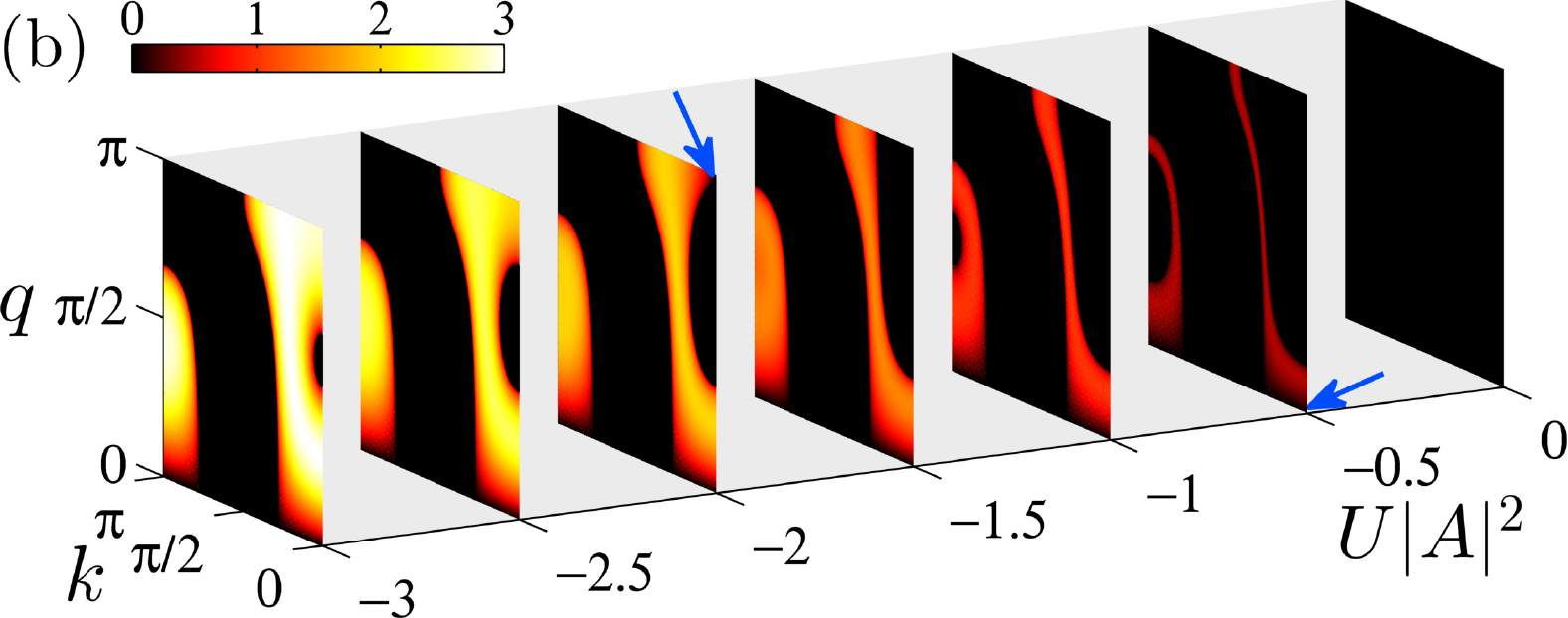}\\
\includegraphics[height=2.99cm]{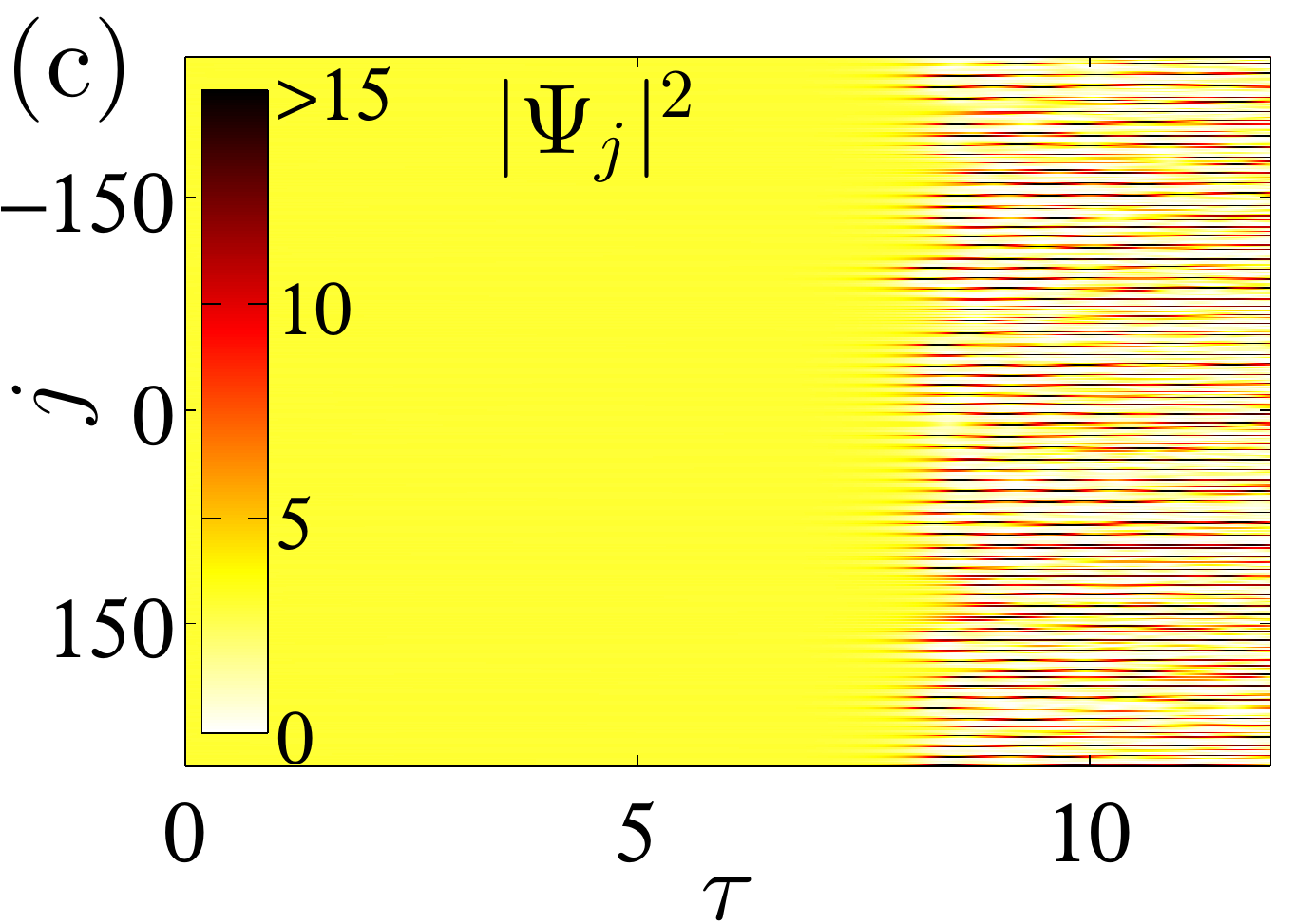} \includegraphics[height=2.99cm]{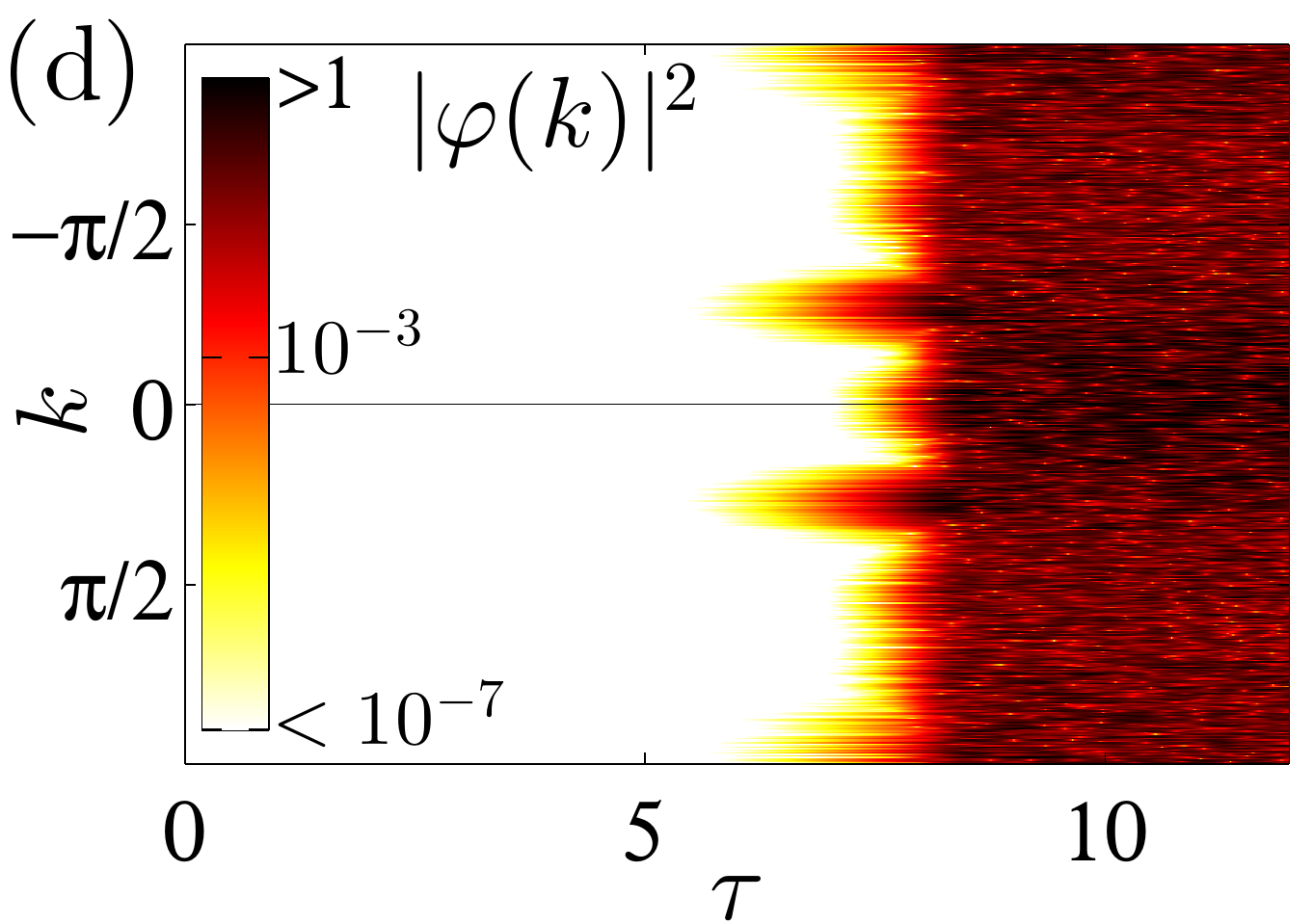} \,
\caption{(Color online) (a,b) Absolute value of the imaginary part of the linearization eigenvalue $\omega_\pm$ as a function of the unperturbed wave number $k$, the perturbation $q$ and the nonlinearity $U|A|^2$. Parameters are $N=2, t_1=1$ and (a) $t_N=0$, (b) $t_N=1$. In the latter case, if $U|A|^2 < -2$, the wave at $k=0$ exhibits two disconnected (in $q$-space) intervals of modulational instability, as indicated by the arrows. 
(c,d) Decay of a plane wave with $k=0$ and $|A|^2=3$, at $t_N=t_1=-U=1$: evolution of the density in direct space (c) and the normalized discrete Fourier transform of $\Psi_j$ (d).\label{fig:mi}}\end{figure}
 
For any amplitude $A$ and any quasi-momentum $k$, the plane wave $\psi_j = A \exp(ikj)$ is a stationary solution to Eq.~(\ref{eq:sdnls}),
but its stability properties may be drastically different for different parameters.
To probe this, one introduces the perturbation ansatz
\begin{equation}
 \Psi_j = \left[ A + \varepsilon a e^{i(qj - \omega \tau)} + \varepsilon b^* e^{-i(qj - \omega^* \tau)} \right] e^{i[kj-\mu(k) \tau]},
\end{equation}
inserts it into Eq.~(\ref{eq:dnls}) and linearizes in $\varepsilon$, which yields an eigenvalue problem for the perturbation frequency $\omega$
with the pair of solutions \cite{Hennig2001}
\begin{equation}
 \omega_\pm = 2 \sum_{l=1,N} t_l \sin ( q l) \sin(k l) \pm 2 \sqrt{ F (F+ U |A|^2)}.
\label{eq:om}
\end{equation}
Here, $F=F(k,q)=2 \sum_l t_l \cos (k l) \sin^2 \frac{q l}{2}$. In the limit of only NN hopping, Eq.~(\ref{eq:om}) reduces to the corresponding result of \cite{Smerzi2002}.
Now modulational instability of a plane wave of amplitude $A$ and wave number $k$ towards a perturbation of wave number $q$ is signalled by an imaginary part of the corresponding $\omega_\pm$, which arises from the square root 
in Eq.~(\ref{eq:om}) if $F$ and $F+U|A|^2$ are of opposite sign.

Figs.~\ref{fig:mi}(a,b) display $| \text{Im} \, \omega_\pm |$ as a function of $k$, $q$ and amplitude $|A|^2$ for a NN model ($t_N=0$) and a zigzag model with $N=2$ and $t_N=1$, respectively. 
In both cases, the nonlinearity is chosen as attractive. Since $F$ is even in $k$ and $q$, it is sufficient to consider positive quasi-momenta here.
The first apparent feature in Figs.~\ref{fig:mi}(a,b) is that in all cases instabilities towards small perturbation wave numbers $q$ show up at small amplitudes already, 
and the intervals of small-$q$ instability drastically change in the presence of second-neighbor hopping.
This is immediately linked to the deformation of the linear dispersion relation discussed before, cf. Fig.~\ref{fig:setup}(b).
To see this, expand $F$ to second order in $q$ which yields
\begin{equation}
 F(k,q\ll 1) \approx \frac{q^2}{2} \left(t_1 \cos k + t_N N^2 \cos kN \right) = \frac{q^2}{4} \mu_0''(k).
\end{equation}
Thus, if $q$ is small and the curvature of the dispersion is positive, $\mu_0''(k) > 0$, then $F>0$.
Correspondingly, MI requires $F+U|A|^2<0$ which is possible only for $U<0$; in that case, however, an infinitesimal amplitude $A$ is sufficient for infinitesimal-$q$ instability, since $F \propto q^2$.
In contrast, if $\mu_0''(k) < 0$, then small-$q$ instability arises at small amplitudes if $U>0$.
In other words, regions of the Brillouin zone in which the curvature is positive (negative) are prone to modulational instability at small $q$ if and only if the 
nonlinearity is attractive (repulsive).
In the NN model, this is already the essence of the MI analysis: The dispersion curvature switches from positive to negative at $k=\pi/2$, and
for the attractive (repulsive) DNLS plane-wave instability occurs for $k<\pi/2$ ($k>\pi/2$) if $t_N=0$, see Fig.~\ref{fig:mi}(a).
Increasing the amplitude further only continuously widens the range of $q$-values towards which there is MI.

In contrast, with the $N$-th neighbor hopping present, the MI phenomenology is a lot richer. The small-$q$ instability is still captured by the dispersion curvature,
but for the parameter values underlying Fig.~\ref{fig:mi}(b), $N=2$ and $t_N=1$, the $\mu_0(k)$ curve now has two inflection points in the interval $0<k<\pi$, see Fig.~\ref{fig:setup}(b).
Correspondingly, there are three distinct intervals of alternating stability and instability towards small $q$ in the MI diagram.
Furthermore, it can also be observed that certain values of $k$ (e.g. in the vicinity of $k=\pi/2$) are stable towards infinitesimal $q$, but unstable towards larger values of $q$,
which never occurs in the NN model.

A further crucial feature to be noted in Fig.~\ref{fig:mi}(b) is the peculiar structure of the instability intervals at $k=0$ for varying $q$.
Here, according to the previous discussion, instability towards small $q$ sets in at small amplitudes, and increasing $|A|^2$ initially only widens the range of unstable $q$ near $q=0$.
However, when crossing a critical amplitude a second interval of unstable $q$ values emerges from $q=\pi$ as indicated by the arrow in the figure.
This is an example of a generic feature of the lower band-edge wave MI in the helicoidal DNLS equation with attractive nonlinearity. 
To see this, note that 
\begin{eqnarray}
 F(k=0,q)&=&2 t_1 \sin^2 \frac{q}{2} + 2 t_N \sin^2 \frac{q N}{2} \nonumber\\
&=&t_1+t_N +\frac{1}{2} \mu_0(q) \geq 0.
\end{eqnarray}
Thus, modulational instability of the $k=0$ wave requires $F+U|A|^2 < 0$ and can only occur for negative $U$.
For a given $q$, the lower band edge wave becomes unstable when 
\begin{equation}
 |A|^2 > \frac{2t_1+2t_N +\mu_0(q)}{(-2U)}, \quad (U<0).
\label{eq:crampl}
\end{equation}
The critical amplitude is thus controlled by the linear dispersion relation: Values of $q$ with smaller $\mu_0(q)$ become unstable for smaller amplitudes.
The global minimum of $\mu_0(q)$ is at $q=0$, so the first instability interval always emerges from there (at infinitesimal amplitudes, as discussed above). 
But beyond this, every additional local minimum in the dispersion will cause the corresponding $q$ to turn 
unstable at smaller amplitudes than its vicinity, thus inducing a separate $q$-interval of instability.
For $N=2$ and $t_N=1$, the only additional local minimum of $\mu_0(q)$ is located at $q=\pi$, see Fig.~\ref{fig:setup}(b), and the corresponding critical amplitude for its instability is given by
\begin{equation}
 |A|^2 = \frac{2t_1+2t_N +\mu_0(q=\pi)}{(-2U)} = \frac{2t_1}{(-U)},
\end{equation}
in agreement with what is observed in Fig.~\ref{fig:mi}(b).

For general $N$, $t_N$, the minima of $\mu_0(q)$ cannot be obtained in closed form, but unless $t_N$ is small, their values are predominantly determined by the short-wavelength $\cos(kN)$ term in Eq.~(\ref{eq:ldisp}).
Then the dispersion minima can be well approximated by the expression 
\begin{equation}
 q_m = m\frac{2\pi}{N}, \qquad m=0,1, \dots ,  \lfloor \frac{N}{2} \rfloor
\label{eq:qmin}
\end{equation}
Now from Eq.~(\ref{eq:crampl}), the critical amplitude at which the lower band edge wave becomes unstable towards one of these minima $q_m$ is approximately given by
\begin{equation}
 |A|^2_{(m)} =  \frac{2t_1+2t_N +\mu_0(q=q_m)}{(-2U)} = \frac{2t_1}{(-U)} \sin^2 \left( m \frac{\pi}{N}\right).
\label{eq:crampl2}
\end{equation}

Thus, the MI of the lower band-edge wave in the attractive version of Eq.~(\ref{eq:dnls}) is characterized
by a cascading destabilization of disconnected $q$-intervals which subsequently turn unstable with increasing amplitude.
The critical amplitude at which MI towards a new interval of quasi-momenta sets in is approximately given by Eq.~(\ref{eq:crampl2}).

In Figs.~\ref{fig:mi}(c,d) we show an example of the dynamics triggered by seeding a modulationally unstable plane wave with some initial white noise. 
While in direct space the formation of strongly localized (mostly single-site) density maxima is observed, $\varphi(k)$, the 
discrete Fourier transform of $\Psi_j$, reveals a transient predominant rearrangement towards the most unstable $q$-values, cf. the corresponding $U|A|^2=-3$ slice in (b), 
before eventually delocalization over the entire quasi-momentum space sets in.

\section{Localized breather solutions}
\label{sec:br}
The DNLS equation with NN hopping features a variety of intrinsically localized stationary solutions,
which are termed breathers or discrete (bright) solitons in the literature.
We will discuss in this section the impact of the $N$-th neighbor hopping in our model Eq.~(\ref{eq:dnls})
on the properties of such localized solutions, focusing primarily on the on-site breathers, which at 
large nonlinearities are characterized by the predominant occupation of a single lattice site.
For the special case of $N=2$ and attractive nonlinearity, it has been shown in \cite{Efremidis2002} that the $t_N$ hopping
crucially modifies the stability properties of the on-site breather branch,
inducing a frequency interval of instability,
indications of which have also been observed experimentally \cite{Szameit2009}.

Generically, at weak nonlinearities the localized breathers are expected to approach 
the band-edge plane waves of the linear problem \cite{Flach1996}.
In the 1D NN-DNLS equation, the on-site breather delocalizes towards the lower (upper) band-edge wave
for attractive (repulsive) nonlinearity, respectively.
Commonly, the long-wavelength MI of these plane waves is taken to be a hint that localized, breather-type solutions exist.
In \cite{Efremidis2002}, this reasoning was extended to predict that for $N=2$, $U<0$ and sizable $t_N$
localized solutions also bifurcate from the staggered $k=\pi$ plane wave, which forms a 
local minimum of the dispersion $\mu_0(k)$, see Fig.~\ref{fig:setup}(b), and also exhibits small-$q$ MI, see Fig.~\ref{fig:mi}(b).
This was further justified within a continuum approximation, 
and the existence of such unusual staggered breather solutions in the attractive DNLS model with second-neighbor hopping was demonstrated numerically.

In contrast, our discussion of the attractive helicoidal DNLS Eq.~(\ref{eq:dnls}) with $U<0$, mostly focuses on the localized breather solutions that stem from the global minimum of the dispersion, $k=0$.
We argue in the following that the instability interval of this breather branch observed in \cite{Efremidis2002,Szameit2009} for $N=2$ can be linked to the peculiar two-stage MI of the parental lower band-edge wave discussed in Sec.~\ref{sec:MI}.
This argument is then extended to arbitrary $N$ and allows us to give analytical estimates of the parameters at which the breather instability sets in.
Subsequently, we discuss separately the on-site breathers at repulsive nonlinearity, where the situation can be radically different due to the split upper band edge.

\subsection{Attractive nonlinearity}
In this subsection we assume $U<0$.
To obtain the on-site breather modes of Eq.~(\ref{eq:sdnls}) numerically, we start from the so-called anti-continuum limit of $t_1=t_N=0$,
in which case a single-site excitation is an exact solution. Subsequently, $\mu$ is kept fixed while $t_1$, $t_N$ are iteratively tuned to their desired values 
and the breather solution is continued numerically using a Newton solver \cite{Kelley2003}.
In the next step, the breather solution is scanned as a function of $\mu$ at fixed $t_1$, $t_N$, $U$.
A key quantity in characterizing the breather branch is its $\norm(|\mu|)$ curve,
where $\norm=\sum_j |\psi_j|^2$ denotes the norm of the stationary solution.

\begin{figure}[ht!]
 \includegraphics[height=2.99cm]{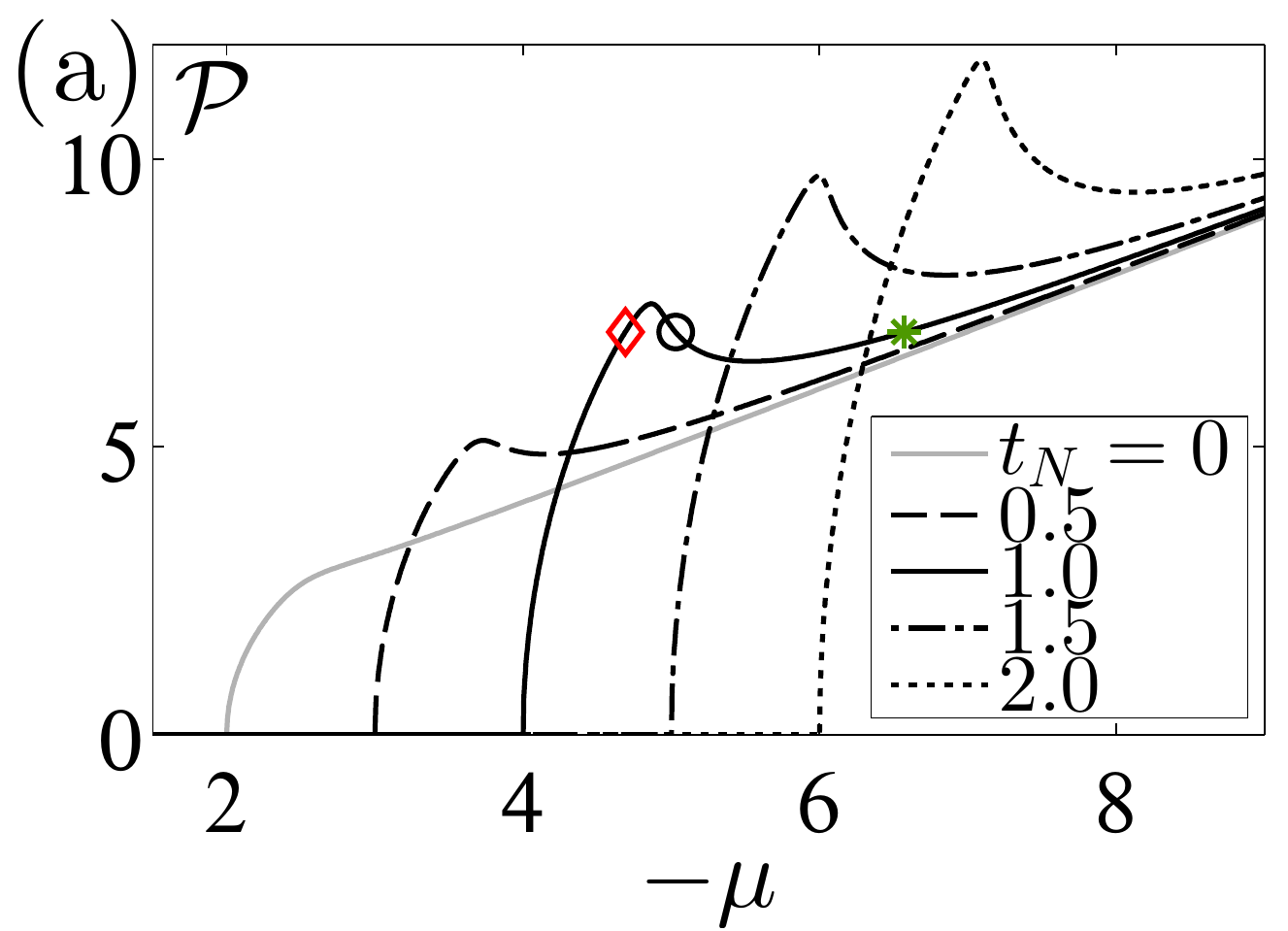} \includegraphics[height=2.99cm]{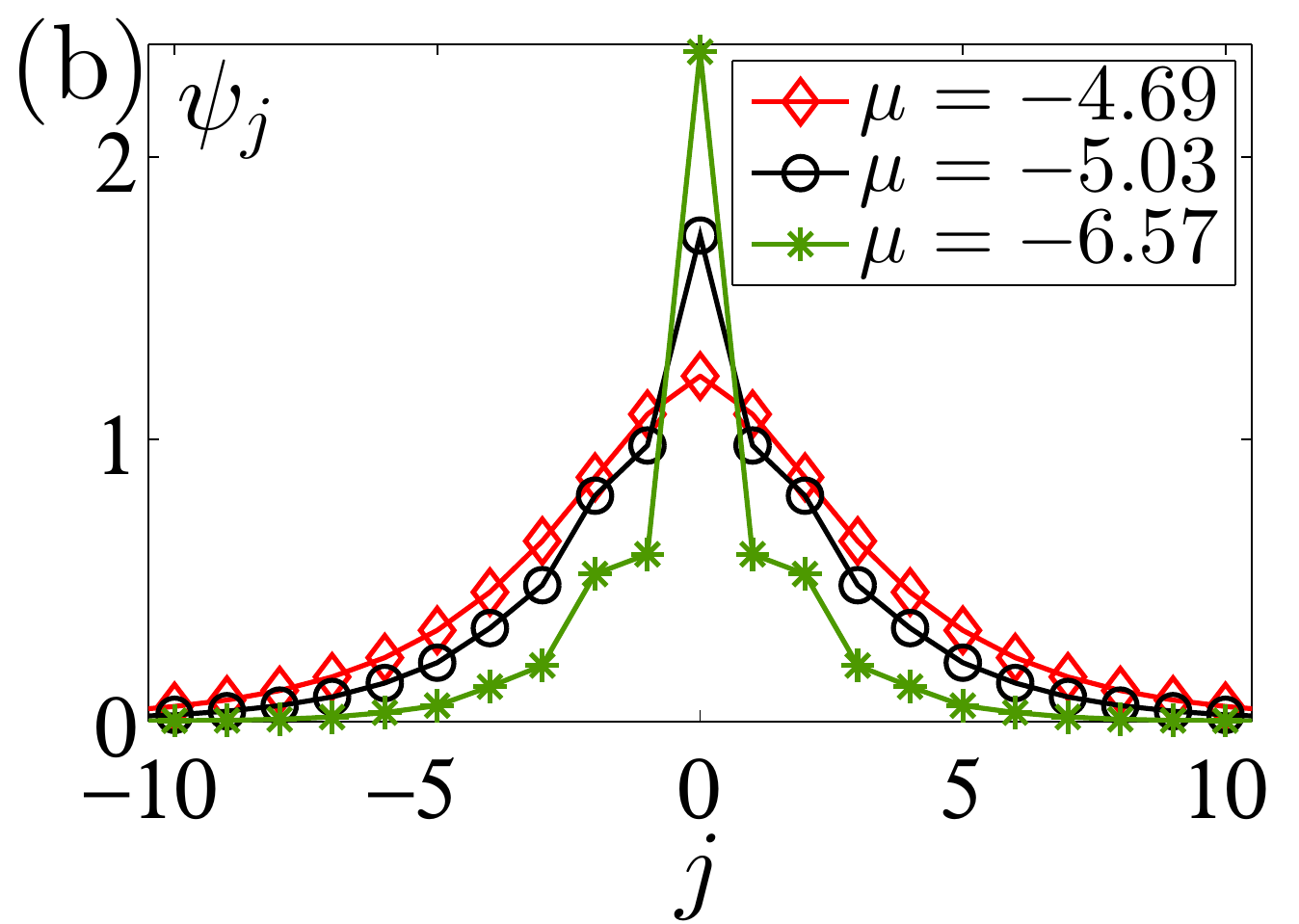}\\
\includegraphics[height=2.99cm]{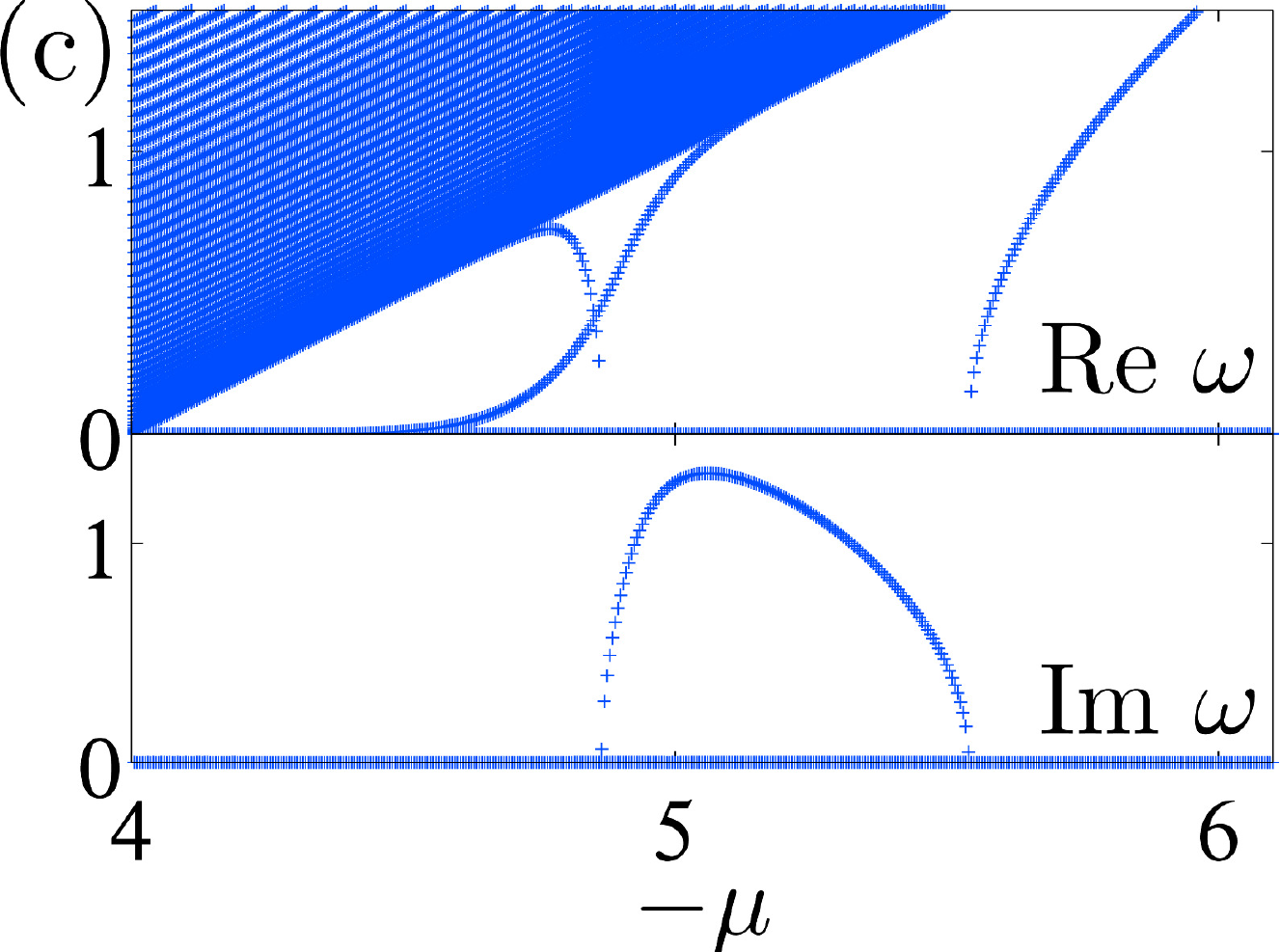} \includegraphics[height=2.99cm]{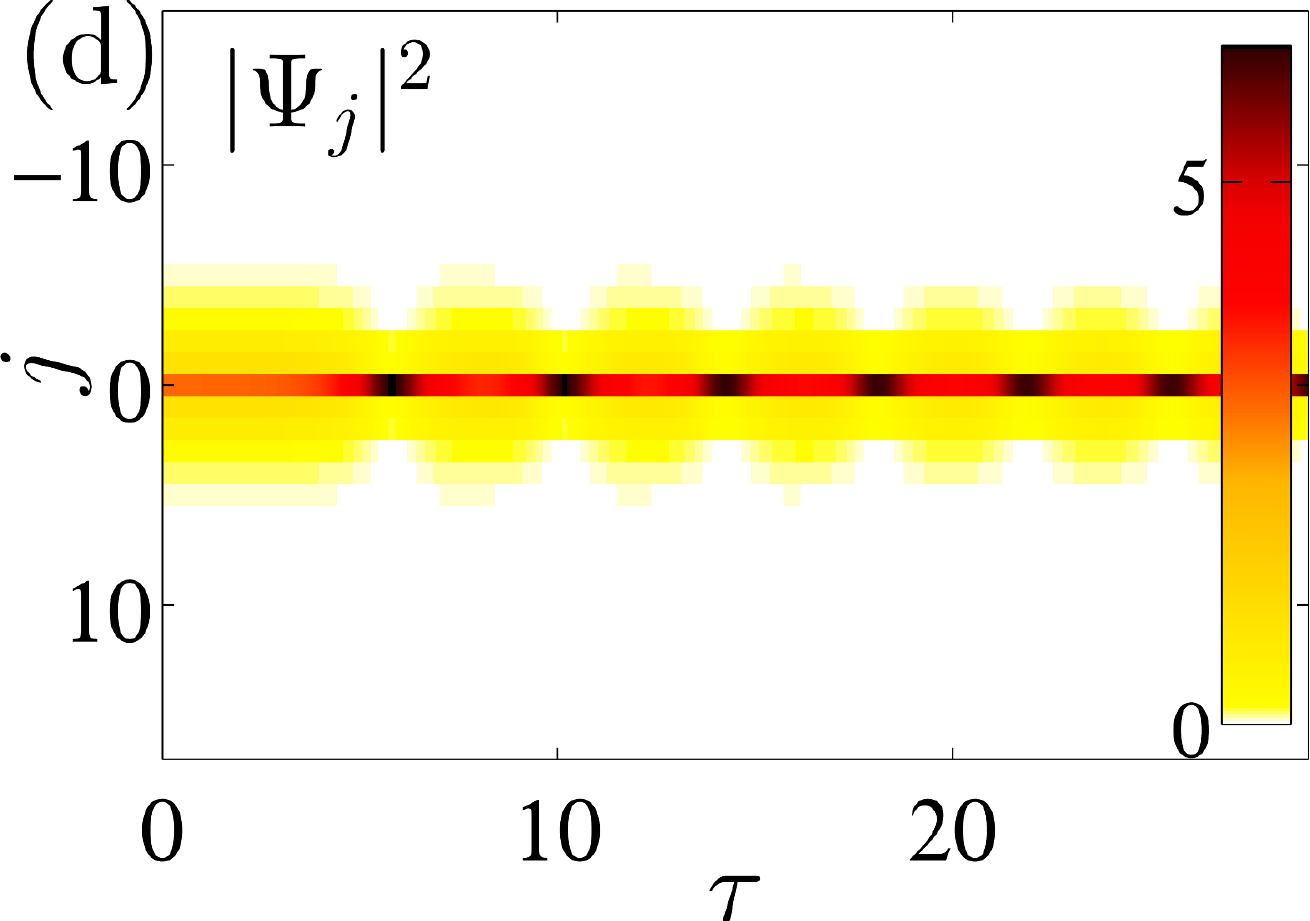} \\
\includegraphics[height=2.99cm]{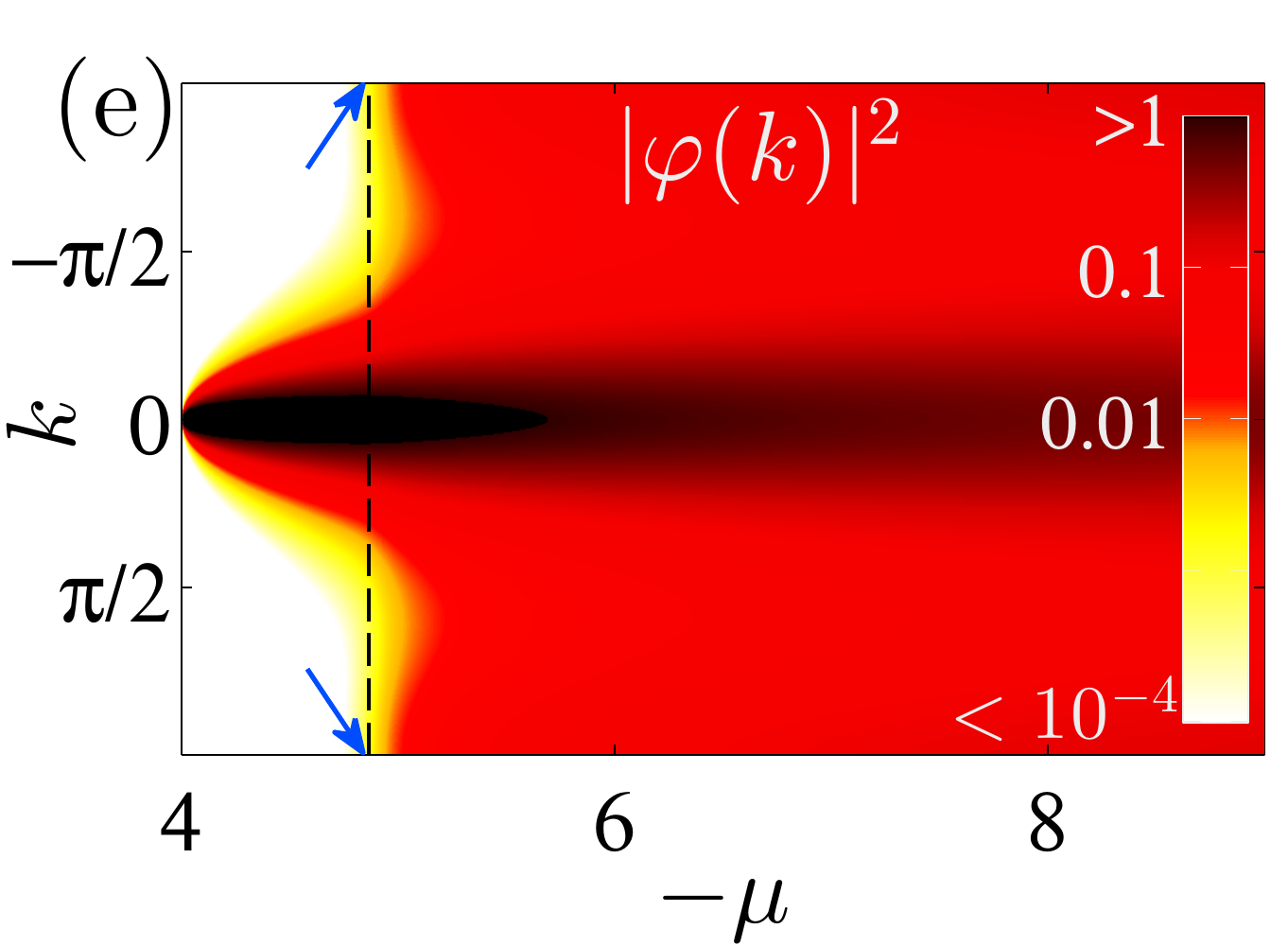} \includegraphics[height=2.99cm]{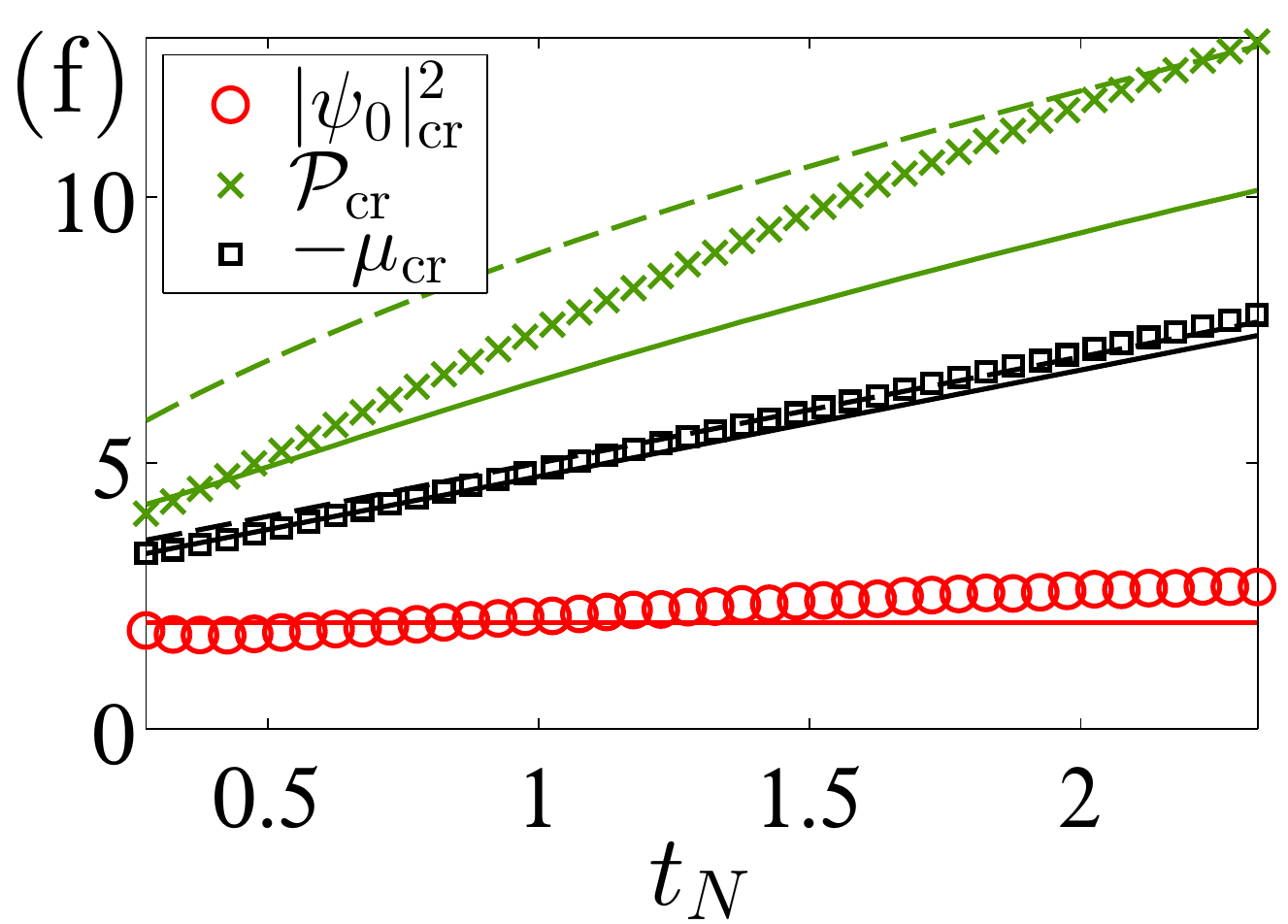} \,
\caption{\label{fig:N2}(Color online) (a) $\norm(\mu)$ curves for the on-site breather branch of solutions at $N=2$ and different $t_N$. (b) Direct space profiles at $t_N=1$ and values of $\mu$ as indicated by the markers in (a).
 (c) Linearization spectrum as a function of frequency $\mu$ of the branch of solutions at $t_N=1$.
The full spectrum is symmetric under $\omega \rightarrow -\omega$, $\omega \rightarrow \omega^*$, so
we restrict to the first quadrant of the complex plane.  (d) Decay dynamics of the unstable mode at $\mu=-5.03$ [circular black markers in (b)] seeded with white noise.
(e) Normalized discrete Fourier transform of the branch of solutions for $t_N=1$ and varying $\mu$. The dashed line denotes the position at which $\norm(|\mu|)$ has its local maximum.
(f) Frequencies, norms and peak amplitudes of the breather solutions at the respective local maximum of $\norm(|\mu|)$ as a function of $t_N$. Markers show the numerical results, 
solid lines for $\mu_{\rm cr}$, $\norm_{\rm cr}$ are obtained from Eqs.~(\ref{eq:app2a}), (\ref{eq:app2b}), dashed lines from Eqs.~(\ref{eq:app1a}), (\ref{eq:app1b}). 
The solid red line depicts $|A|^2_{(1)}$ as predicted by Eq.~(\ref{eq:crampl2}). Throughout $t_1=1$, $U=-1$.}\end{figure}

Let us first discuss the case of $N=2$, which has been considered before in \cite{Efremidis2002,Szameit2009}.
The numerically obtained $\norm(|\mu|)$ curves of the on-site breather for varying values of $t_N$ are shown in Fig.~\ref{fig:N2}(a).
While in the NN model ($t_N=0$) the norm $\norm$ is a monotonically increasing function of $|\mu|$, for $t_N \gtrsim 0.26$ it exhibits a local maximum and a minimum,
such that with increasing $|\mu|$ the slope changes from positive to negative and then to positive again.
Evidently, the breather branches start to exist at $\mu = \mu_0(k=0)=-2(t_1+t_N)$ with zero norm, asymptoting to the lower band-edge wave in this linear limit.
A selection of breather profiles at $t_N=1$ is shown in Fig.~\ref{fig:N2}(b), for values of $\mu$ as given in the legend and indicated by the markers in Fig.~\ref{fig:N2}(a).
It can be seen here that with increasing $|\mu|$ (away from the lower band edge) the breather localizes more and more to its central site.
In the interval of negative slope of the $\norm(|\mu|)$ curve, the breather solutions are linearly unstable,
as can be seen explicitly in the frequency-dependent linearization spectrum shown in Fig.~\ref{fig:N2}(c) for the branch of solutions at $t_N=1$.
At the maximum of $\norm(|\mu|)$, one real eigenmode coming from the phonon band crosses to the imaginary axis and remains there, signaling exponential instability, until the minimum of $\norm(|\mu|)$ is reached.
Fig.~\ref{fig:N2}(d) provides an example of the decay dynamics in the unstable interval. Seeding the unstable breather mode with weak white noise leads to its rearrangement into one of the two coexisting stable solutions at 
the same (or, given some radiation loss, a slightly smaller) norm. In the specific run shown in Fig.~\ref{fig:N2}(d), the final state oscillates close to the $\mu=-6.57$ solution of Fig.~\ref{fig:N2}(b),
but for other noise realizations we also encounter cases of decay towards the $\mu=-4.69$ solution.
Apart from the emerging unstable mode, the linearization spectrum in Fig.~\ref{fig:N2}(c) also features a zero mode rising towards the phonon band for increasing $|\mu|$,
reflecting the loss of translational invariance with increasing nonlinearity \cite{Kevrekidis2009}.

A remarkable feature of the breather rearrangement towards a single-site excitation with increasing frequency is revealed by a study of the discrete Fourier transform $\varphi(k)$
of the breather profile, see Fig.~\ref{fig:N2}(e). 
Here it is seen that near the linear limit the breather is localized in the vicinity of $k=0$ in quasi-momentum space, and 
for small $|\mu|$ the increasing direct-space localization is accompanied by a broadening of the Fourier peak at $k=0$.
However, coinciding with the critical frequency $\mu_{\rm cr}$ at which the $\norm(|\mu|)$ curve has its maximum, $\varphi(k)$ changes drastically, picking up large contributions first near $k=\pm \pi$ 
which subsequently quickly extend into the full quasi-momentum space.

This observation, and in particular the prominent role of the $k=\pm \pi$ quasi-momenta suggest an interpretation linking to our above discussion of the $k=0$ plane-wave MI.
Close to the linear limit, at small amplitudes, the breather branch is highly delocalized and resembles the $k=0$ band edge wave, but with a wide amplitude envelope.
Despite this similarity, however, the breather is not affected by the long-wavelength, small-$q$ MI experienced by the parental $k=0$ wave. 
The localizing envelope of the breather may be thought to have just the right shape to prevent modulational-type instability towards the growing interval of unstable $q$ of the underlying plane wave,
and make the breather linearly stable. 
When viewed in Fourier space, the breather envelope is generated by a population of the small-$q$ momenta causing the plane-wave MI. 
Now in the NN-DNLS model, the interval of unstable $q$ of the lower band edge wave continuously extends away from zero with increasing nonlinearity, see Fig.~\ref{fig:mi}(a),
and the breather envelope continuously deforms and remains stable.
In the presence of the $t_N$-hopping, however, we have pointed to the fact that upon crossing a finite critical amplitude the MI of the $k=0$ wave changes qualitatively,
in that beyond this point a new instability arises from $q=\pi$ for $N=2$ (or, more generally, from the second-lowest band minimum near $q_1 = 2\pi/N$), see Fig.~\ref{fig:mi}(b).
We suggest now to view the destabilization and radical recomposition in $k$-space of the breather branch at the maximum of its $\norm(|\mu|)$ curve 
as a remnant of this second stage of the band-edge wave MI.
In other words, in this picture the localizing envelope of the breather branch is thought to provide the right shape to immunize it 
against the small-$q$ MI of the $k=0$ plane wave at small amplitudes (as in the NN model), but at larger amplitudes where the qualitatively different $q=\pi$ instability
of the $k=0$ plane wave sets in, the breather also picks up an instability.
Subsequently, the breather profile strongly rearranges, evidenced by a sudden spread in quasi-momentum space as seen in Fig.~\ref{fig:N2}(e),
incorporating in particular quasi-momenta in the vicinity of the unstable $q_1$.

To check this idea, let us go back to Eq.~(\ref{eq:crampl2}), which predicts the critical amplitude at which the second $q$-instability interval of the $k=0$ plane wave comes into existence.
This is given by $|A|^2_{(1)} = 2 t_1/(-U) \sin^2 (\pi/N) = 2t_1/(-U)$ for $N=2$. Remarkably, this value does not depend on $t_N$. 
Now if indeed the instability interval of the breather branch relates to the two-stage MI of the $k=0$ plane wave,
one should expect that this instability sets in at a breather amplitude close to the MI critical amplitude of the plane wave, irrespectively of $t_N$.
Since the onset of instability coincides with the critical point where $\norm(|\mu|)$ has its maximum,
the breather amplitude at this critical point should lie in the vicinity of $|A|^2_{(1)}$.
Fig.~\ref{fig:N2}(f) shows a comparison of the numerically obtained breather peak amplitude $|\psi_0|^2_{\rm cr}$ at the critical point as a function of $t_N$,
comparing it to the $t_N$-independent value of $|A|^2_{(1)} = 2 t_1/(-U)$. 
Indeed, although $|\psi_0|^2_{\rm cr}$ is not fully independent of the second-neighbor hopping $t_N$, the agreement with $|A|^2_{(1)}$ is quite good over a wide range of values.
Full quantitative agreement may certainly not be expected here, since obviously the breather solution is distinct from a flat plane wave,
and the above reasoning essentially relies on identifying the localized breather with a plane wave of its central (peak) amplitude.
Since $|\psi_0|^2$ is the largest amplitude of the inhomogeneous breather profile, one may expect that it tends to overshoot $|A|^2_{(1)}$,
and in Fig.~\ref{fig:N2}(f) this is indeed the case in the region of large $t_N$ where the deviations are largest 
(whereas at small $t_N$, where the two-stage MI of the band-edge plane wave starts to wash out, there is a slight trend in the opposite direction).

Going one step further, we can employ the approximate insight into the critical breather amplitude 
to obtain estimates for the critical frequency and norm, respectively.
We employ two alternative (and, to a certain extent, complementary) methods for this, both relying on an 
approximate description of the breather profile for varying $\mu$ near the linear limit,
as described in appendix \ref{app}.
These yield approximate expressions for the breather peak amplitude and norm as a function of the frequency $\mu$.
Then, by the above reasoning, the critical point is reached when the squared peak amplitude reaches $|\psi_0|^2 = |A|^2_{(1)} = 2 t_1 \sin^2 (\pi/N)/(-U)$, from which $\mu_{\rm cr}$ and $\norm_{\rm cr}$ are deduced.
The details of this procedure are given in appendix \ref{app}, and we only provide the results here. 
Using a continuum nonlinear Schrödinger approximation as in \cite{Efremidis2002}, we find
\begin{eqnarray}
 \mu_{\rm cr} &\approx& -2(t_1+t_N) -t_1 \sin^2 \frac{\pi}{N}, \label{eq:app1a}\\
 \norm_{\rm cr} &\approx& \frac{4 t_1}{-U} \sqrt{1 + N^2 \frac{t_N}{t_1}}\sin \frac{\pi}{N}.
\label{eq:app1b}
\end{eqnarray}
On the other hand, we have seen above that below the critical point the breather is localized near $k=0$ in quasi-momentum space
and its direct-space profile exhibits no modulation on the length scale of $N$, 
such that for these near-linear frequencies it can be approximated also by a variational ansatz of exponential shape.
This results in expressions slightly different from the continuum approximation:
\begin{eqnarray}
 \mu_{\rm cr} &\approx& -2(t_1+t_N) -\frac{3}{4}t_1 \sin^2 \frac{\pi}{N} \label{eq:app2a},\\
 \norm_{\rm cr} &\approx& \frac{4 t_1}{-U} \sqrt{1 + N^2 \frac{t_N}{t_1}}\sin \frac{\pi}{N} \nonumber \\
 &&+2N(N^2-1) \frac{t_1}{U} \frac{t_N}{t_1+N^2 t_N} \sin^2 \frac{\pi}{N}.
\label{eq:app2b}
\end{eqnarray}
Fig.~\ref{fig:N2}(f) compares the predictions of Eqs.~(\ref{eq:app1a})-(\ref{eq:app2b}) to the numerically obtained maxima of the $\norm(|\mu|)$ curves,
showing good qualitative (and to some degree, especially for $\mu_{\rm cr}$, also quantitative) agreement.

\begin{figure}[ht!]
 \includegraphics[height=2.99cm]{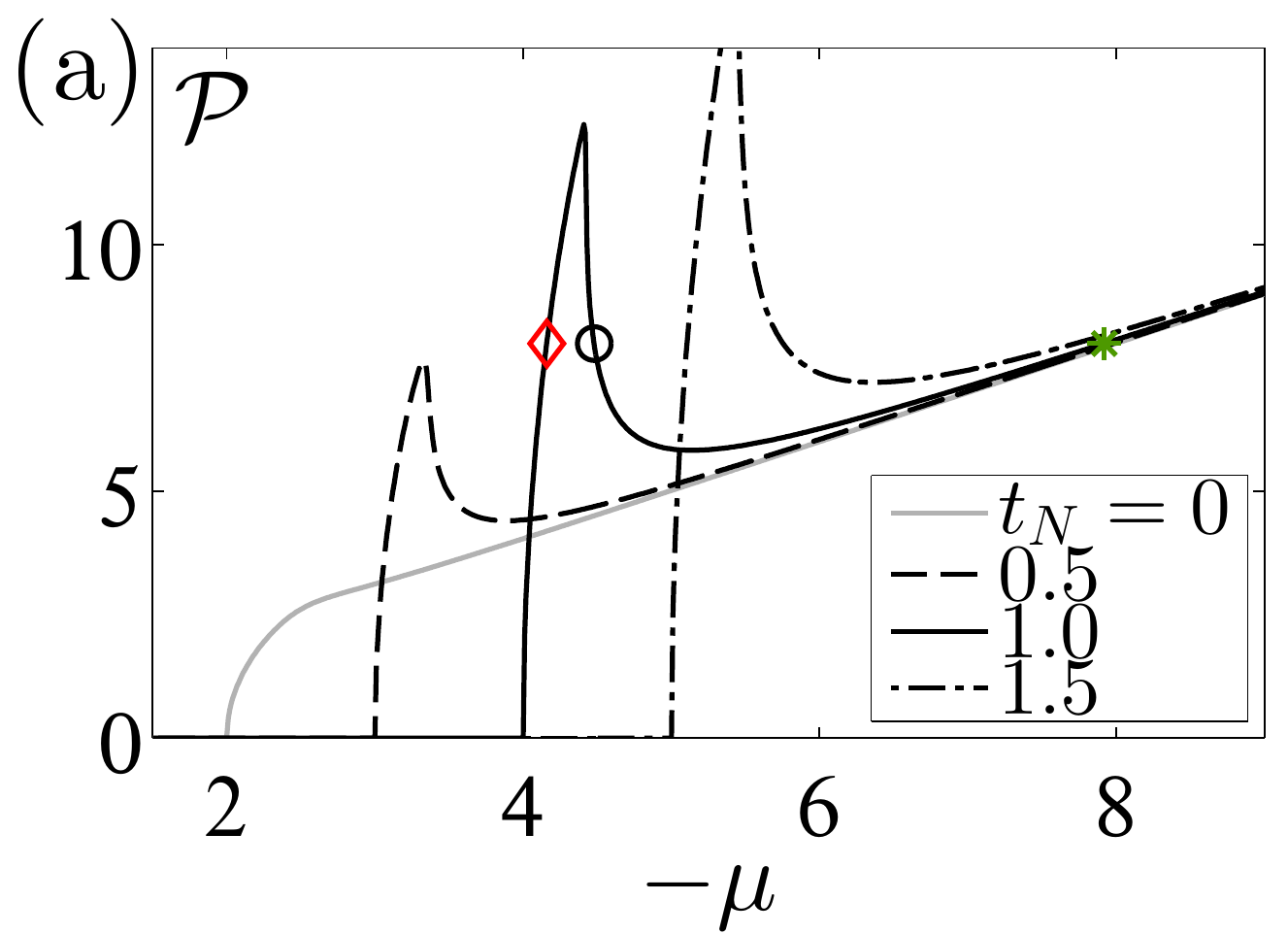} \includegraphics[height=2.99cm]{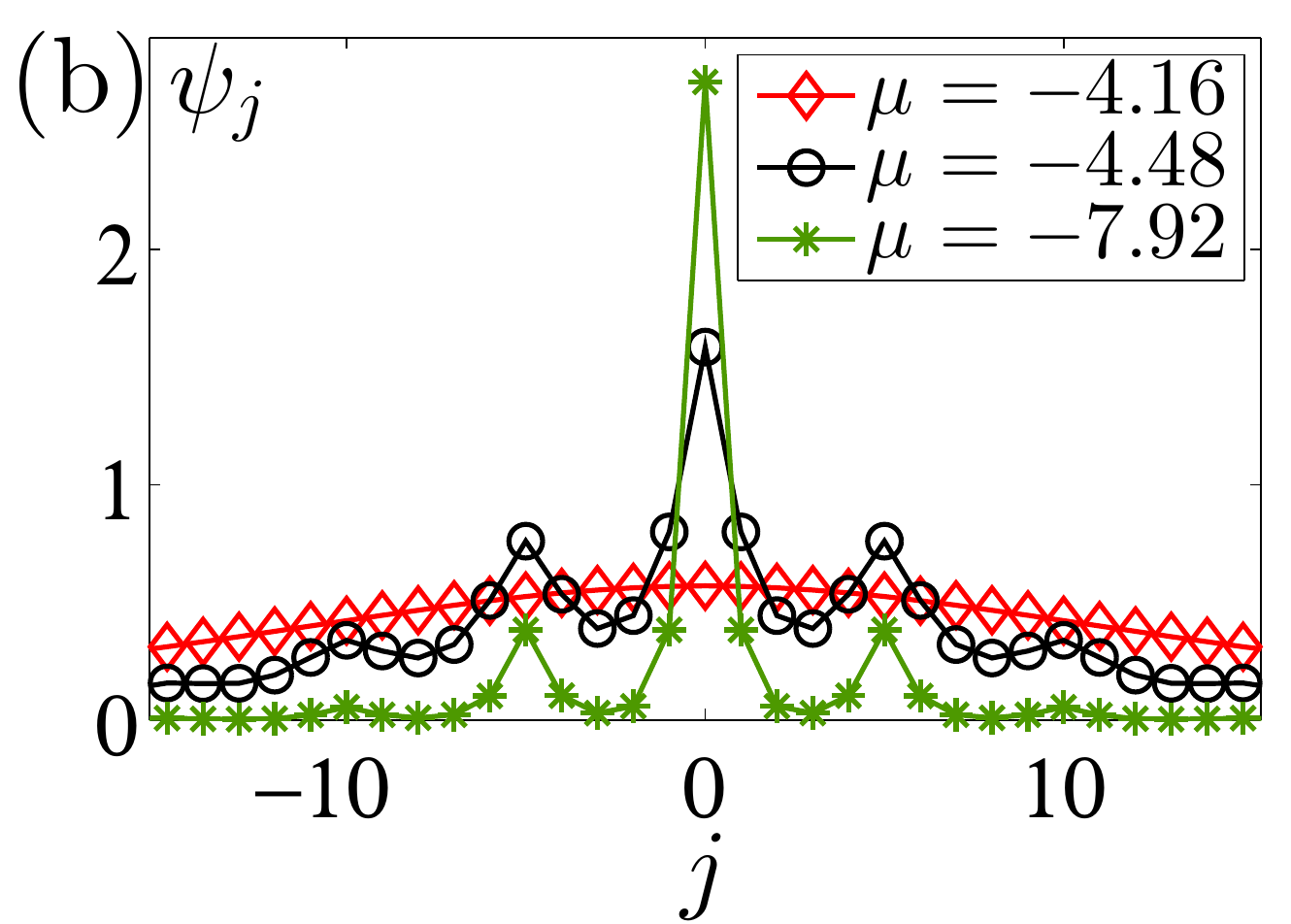}\\
\includegraphics[height=2.99cm]{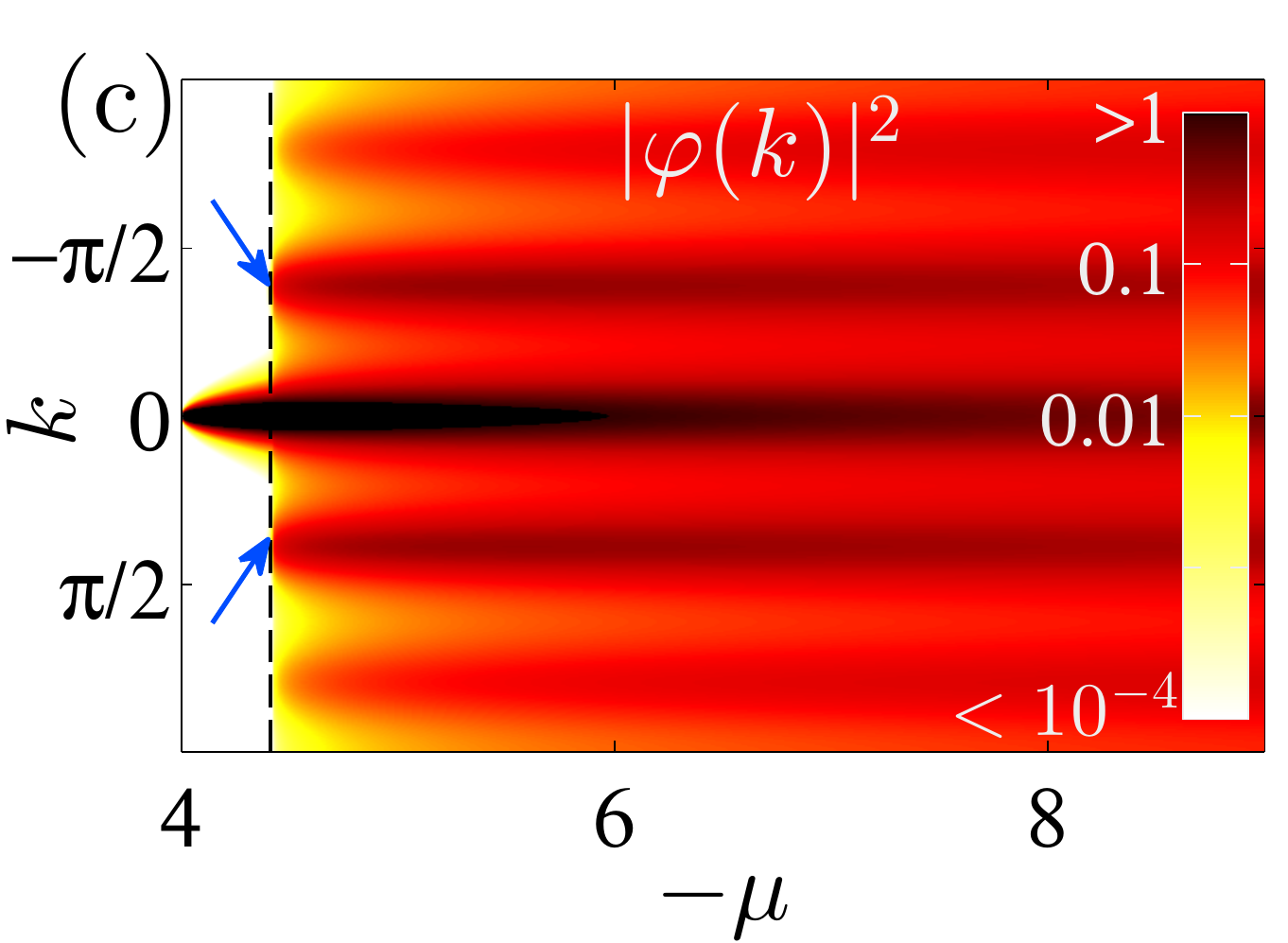} \includegraphics[height=2.99cm]{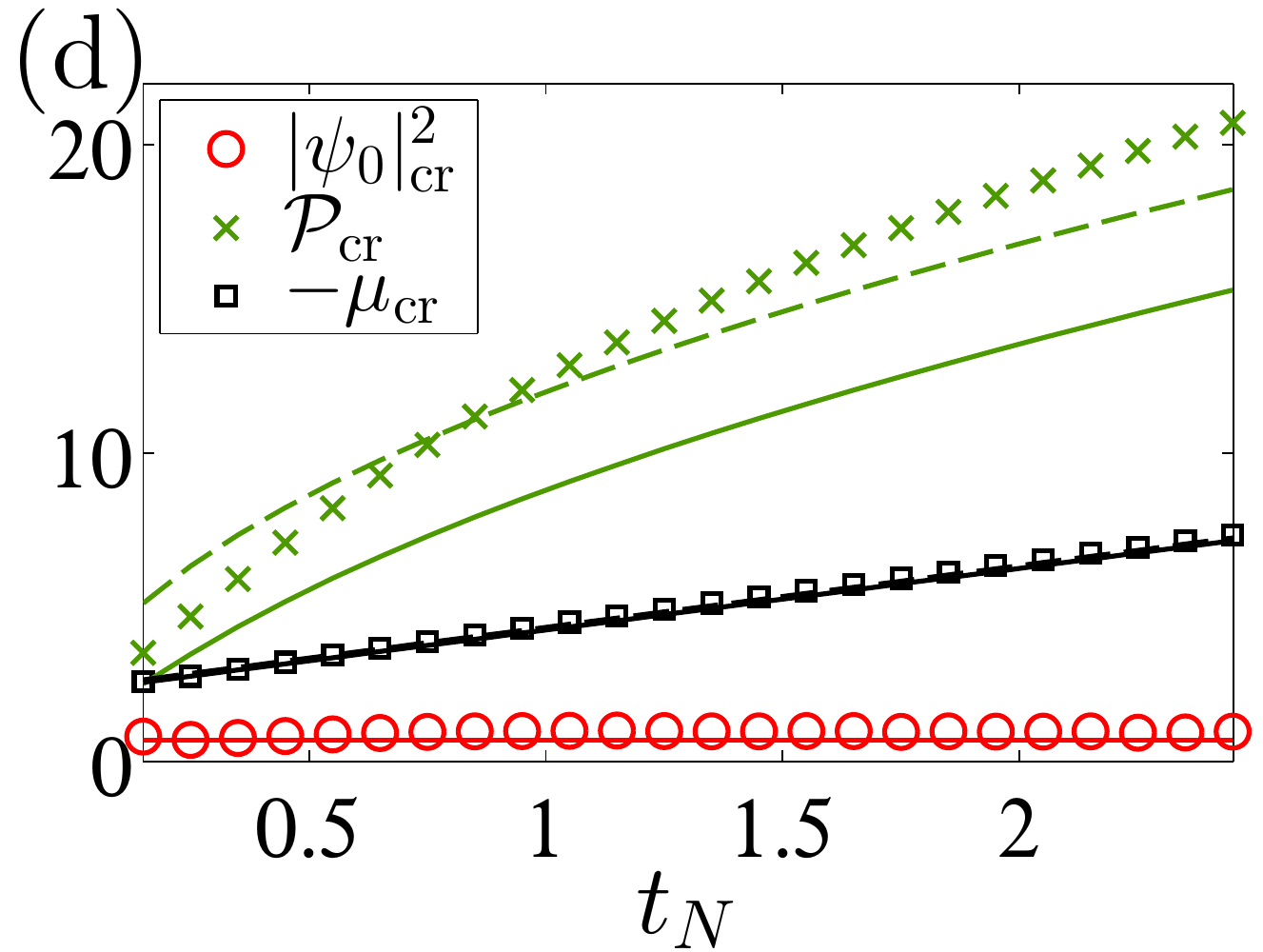} \, \\
  \vspace{2mm}\includegraphics[width=0.45\textwidth]{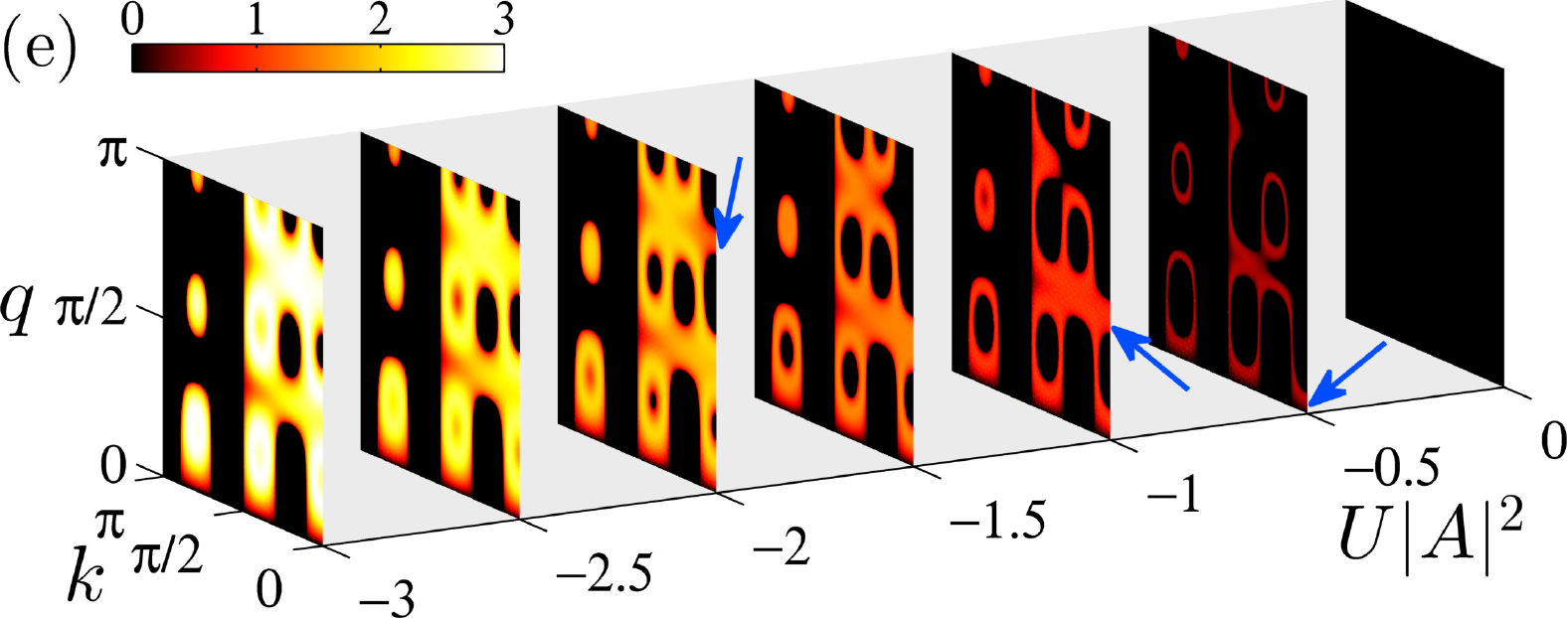}
\caption{\label{fig:N5}(Color online) (a) $\norm(|\mu|)$ curves for the breather branch of solutions at $N=5$ and different $t_N$. (b) Direct space profiles at $t_N=1$ and values of $\mu$ as indicated by the markers in (a).
(c) Normalized discrete Fourier transform of the branch of solutions for $t_N=1$ and varying $\mu$. The dashed line denotes the value of $\mu_{\rm cr}$, arrows point to $\pm q_1$ as in Eq.~(\ref{eq:qmin}).
(d) Frequencies, norms and peak amplitudes of the breather solutions at the respective critical points as a function of $t_N$, 
analytical estimates according to Eqs.~(\ref{eq:app1a})-(\ref{eq:app1b}) (dashed lines) and Eqs.~(\ref{eq:app2a})-(\ref{eq:app2b}) (solid lines, for $-\mu_{\rm cr}$ essentially coinciding with the dashed line on the scale of the figure)
and $|A|^2_{(1)}$ from Eq.~(\ref{eq:crampl2}) (solid red line). (e) MI diagram with colors encoding $|\text{Im}\, \omega_\pm|$ from Eq.~(\ref{eq:om}) for $N=5$, $t_N=1$.
In this case, at $k=0$ three $q$-intervals of MI successively emerge with increasing nonlinearity, indicated by the arrows. Throughout $t_1=1$, $U=-1$.}\end{figure}

So far, we have seen that in the case of $N=2$ the two-stage plane-wave MI analysis admits reasonable estimates of the critical point where the on-site breather branch turns unstable.
In order to evaluate the universality of the above ideas,
let us check their applicability in the framework of a different $N$, choosing the example of $N=5$ here.
Fig.~\ref{fig:N5}(a) shows the numerically obtained $\norm(|\mu|)$ curves for the on-site breather in this case.
Beyond a critical value of the $N$-th neighbor hopping ($t_N \gtrsim 0.11$ here), these curves again exhibit the local maximum-minimum structure 
observed also in Fig.~\ref{fig:N2}(a), while the slope in the vicinity of the maximum tends to be larger for $N=5$ than for $N=2$.

Fig.~\ref{fig:N5}(b) collects a selection of breather profiles, taken at the parameters marked in Fig.~\ref{fig:N5}(a).
It can be seen again that between the linear limit and the maximum of the $\norm(|\mu|)$ curve, the wave packet is comparably wide and has a smooth envelope, showing no sign of modulation on the length scale induced by $N=5$.
This changes beyond the critical point, where the breather profile decays non-monotonically away from its central peak.
As in the $N=2$ case, the frequency interval of negative slope of $\norm(|\mu|)$ is accompanied by linear instability.
Going to even larger values of $|\mu|$, the slope changes to positive and the breather branch is stable again, 
while the profile increasingly localizes to its central site.

The Fourier decomposition of the breather profiles, Fig.~\ref{fig:N5}(c), again reveals a sharp transition at the critical frequency.
For subcritical $|\mu|$, the mode is localized near $k=0$ in quasi-momentum space. Additional contributions arise at $\mu_{\rm cr}$,
but now not from $k=\pm \pi$ (as observed before at $N=2$), but chiefly at the pair of quasi-momenta inside the first Brillouin zone indicated by blue arrows in Fig.~\ref{fig:N5}(c).
From the above general discussion, we expect this pair of peaks at the first local minimum of the dispersion curve,
which for $N=5$ no longer lies at $\pi$, but instead at $q = q_1 \approx 2 \pi/5$ according to Eq.~(\ref{eq:qmin}), agreeing well with the numerical findings.
According to Eq.~(\ref{eq:crampl2}), we expect that the $k=0$ plane wave develops an instability towards this $q$ beyond a critical amplitude
given by $|A|^2_{(1)} = 2 t_1/(-U) \sin^2 (\pi/5) \approx 0.69 \, t_1/(-U)$, see
the plane wave MI diagram in Fig.~\ref{fig:N5}(e).

The distinguished peaks emerging near $q_1 = 2 \pi/5$ at the critical frequency in Fig.~\ref{fig:setup}(c)
are the first indicators that our previous MI-type interpretation of the breather instability is still of value at $N=5$.
Indeed, comparing the breather squared amplitude $|\psi_0|^2_{\rm cr}$ at the critical $\mu$ to the 
plane-wave MI amplitude $|A|^2_{(1)} = 2 t_1/(-U) \sin^2 (\pi/5)$ at which the parental $k=0$ plane wave picks up the $q_1$-instability 
again shows reasonable agreement independently of $t_N$, see Fig.~\ref{fig:N5}(d).
It can be seen that $|A|^2_{(1)}$ tends to underestimate the critical peak amplitude $|\psi_0|^2_{\rm cr}$, as discussed above.
Eqs.~(\ref{eq:app1a})-(\ref{eq:app2b}) are also in qualitative (and approximate quantitative) agreement with the numerical data
and, in particular, correctly capture the trend that for $N=5$ the critical norms $\norm_{\rm cr}$ are substantially larger than for $N=2$, while the frequencies
$\mu_{\rm cr}$ shift closer to the linear limit, such that the $\norm(|\mu|)$ curves rise towards their maximum more steeply for $N=5$.

We have tested the applicability of these general arguments for other values of $N$ and found the expected degree of agreement.
The central breather amplitude $|\psi_0|^2$ at the maximum of the $\norm(|\mu|)$ curve is approximately $t_N$-independent and lies in the vicinity of $|A|^2_{(1)} =  2 t_1/(-U) \sin^2 (\pi/N)$.
The $k$-space Fourier decomposition is localized around $k=0$ below the critical point and at $\mu_{\rm cr}$ delocalizes, starting from peaks near $\pm q_1 = \pm 2\pi/N$.
The critical $\mu_{\rm cr}$ and $\norm_{\rm cr}$ are in overall agreement with Eqs.~(\ref{eq:app1a})-(\ref{eq:app2b}), although, expectably, there are quantitative deviations.
Let us briefly note that for $N=3$ and $N=4$ our numerical results indicate that the near-linear, more delocalized and the localized parts of the breather branch tend to slightly detach from each other with increasing $t_N$.
A detailed inspection of this feature is beyond the scope of the present work.

\begin{figure}[ht]
 \includegraphics[height=2.99cm]{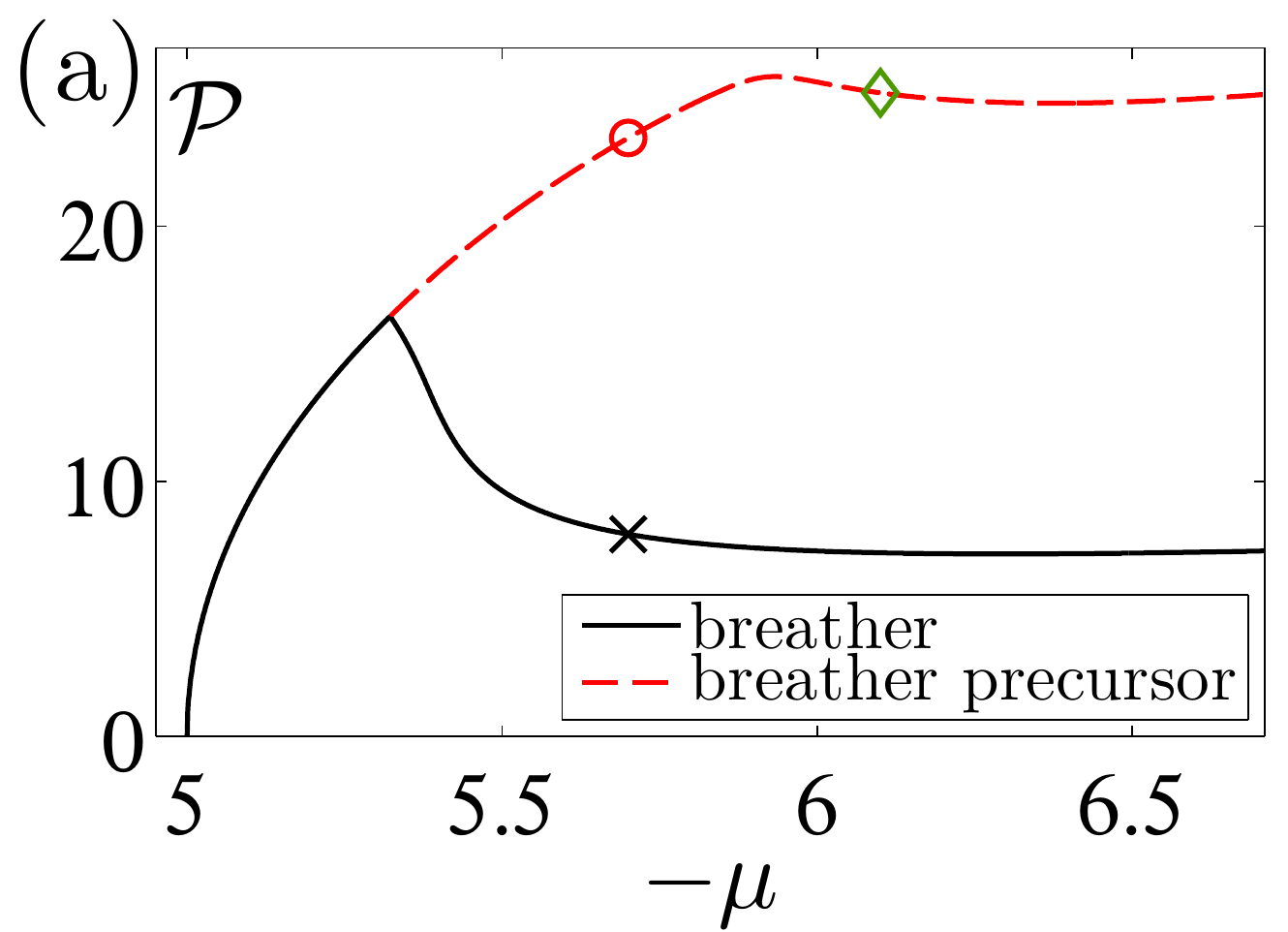} \includegraphics[height=2.99cm]{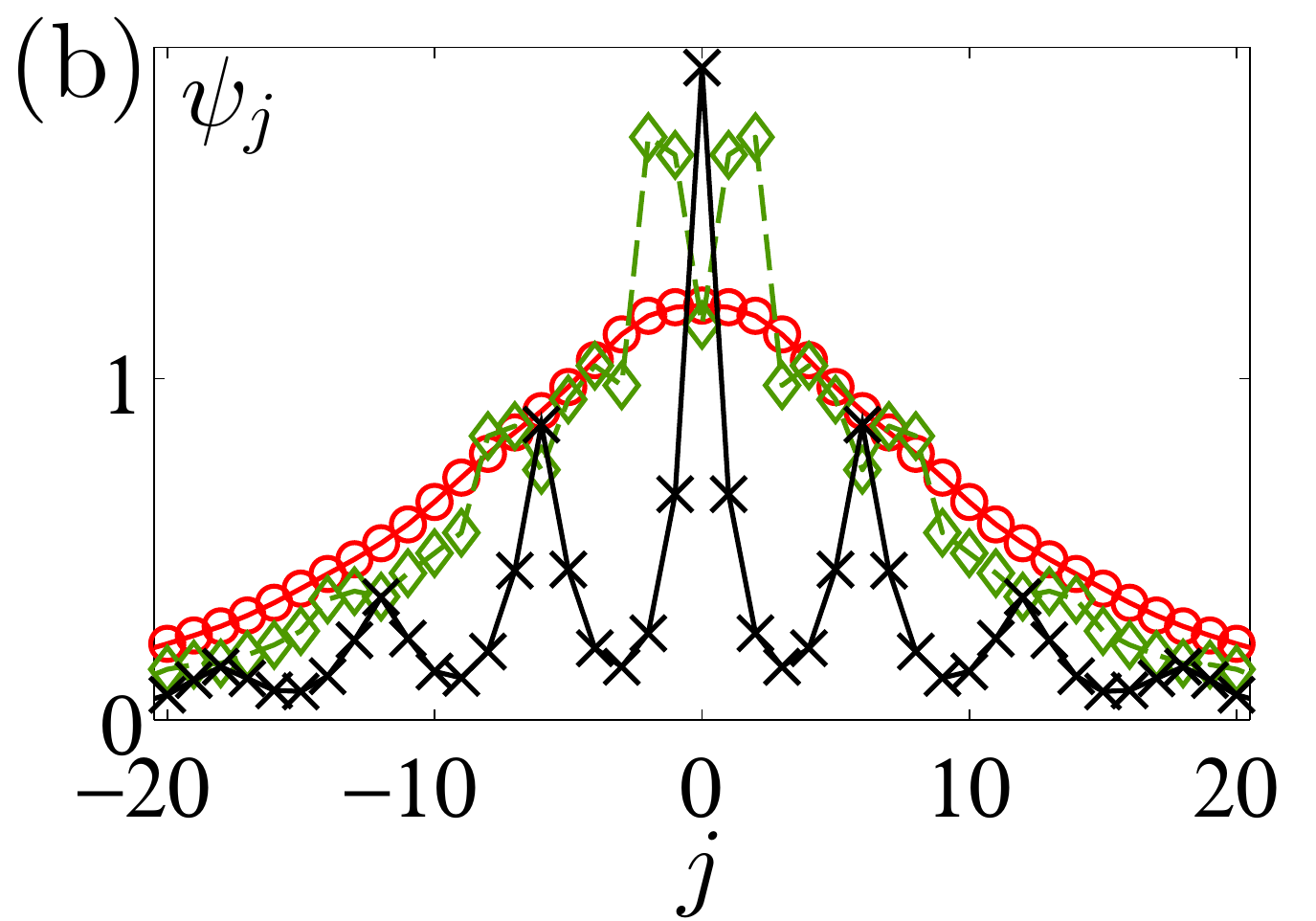}\\
 \includegraphics[height=2.99cm]{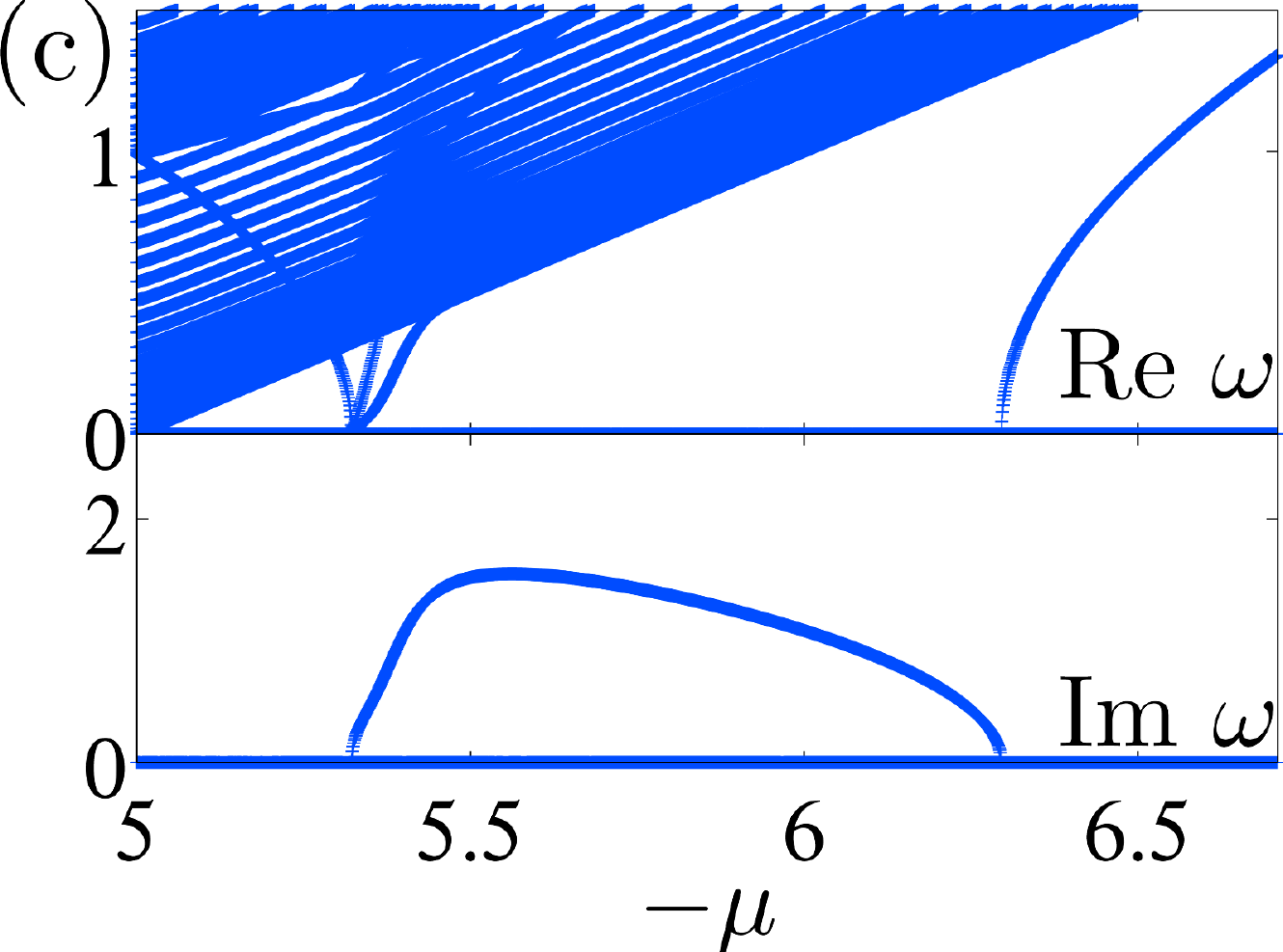} \hspace{1mm} \includegraphics[height=2.99cm]{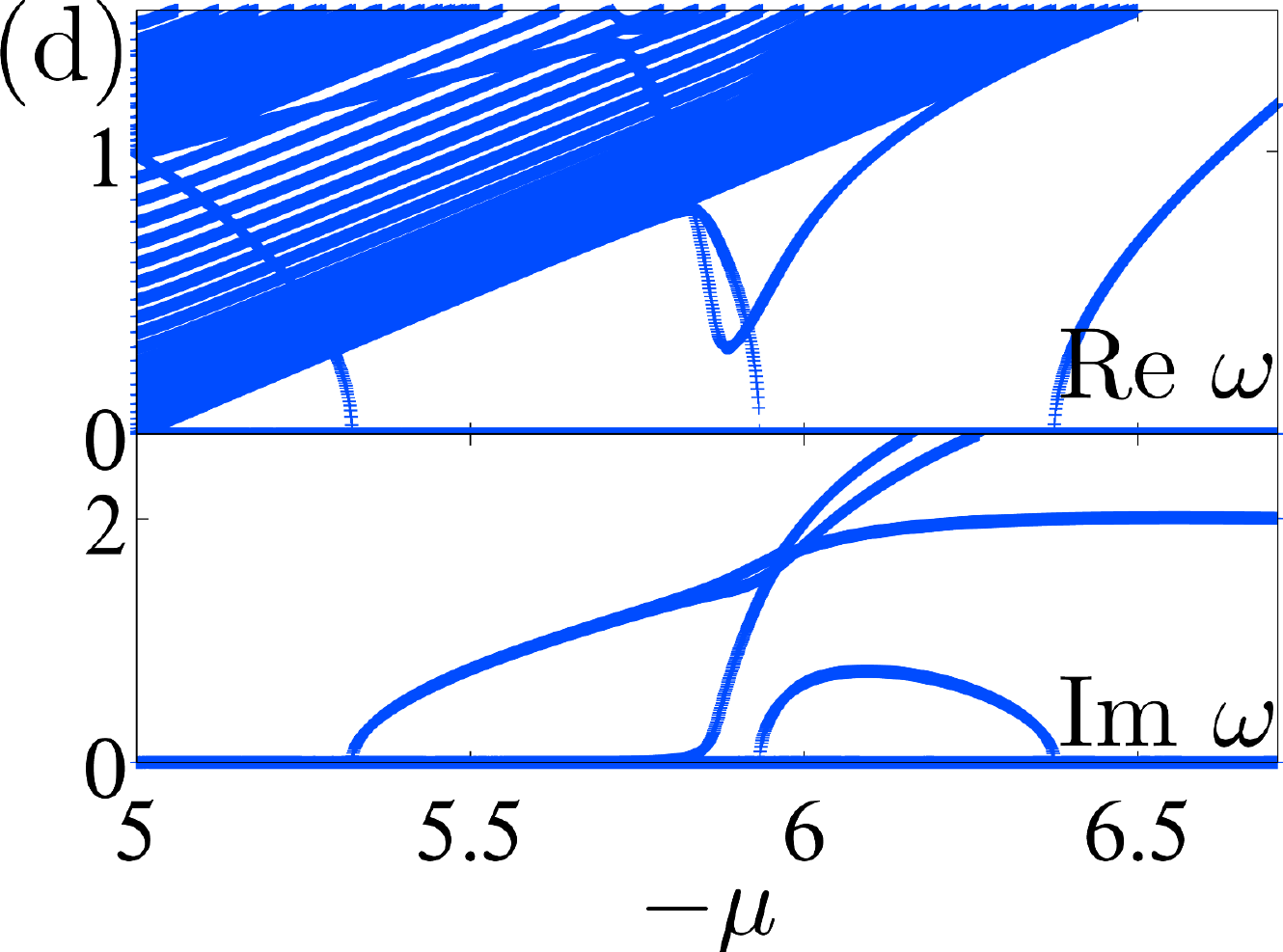}
\caption{\label{fig:N6}(Color online) (a) Norm vs. frequency curves of the on-site breather and the more delocalized precursor-type branch, which continues to exist independently after the breather has passed its maximum norm. 
(b) Selected profiles from the two branches at the parameter values indicated by the respective markers in (a).
(c,d) Linearization spectra as a function of frequency for the breather (c) and the breather precursor (d) branch.
Throughout, $N=6$, $t_N=1.5$, $t_1=-U=1$.}\end{figure}

The relevance of the plane-wave MI analysis for the understanding of localized solutions in the helicoidal DNLS model is further supported by the following observation.
Comparing the $\norm(|\mu|)$ curves of Fig.~\ref{fig:N2}(a) and Fig.~\ref{fig:N5}(a), respectively, already suggests that with increasing $N$, $t_N$ the maxima may tend to assume a cusp-like structure.
Indeed, going beyond $N=5$ we have encountered cases where this cusp shape is very pronounced and it is in fact possible to numerically continue the near-linear part of the breather branch 
(which emerges from the linear limit and is localized near $k=0$ in quasi-momentum space) towards increasing $|\mu|$ beyond the critical value.
An example of this (for $N=6$, $t_N=1.5$) is shown in Fig.~\ref{fig:N6}(a).
In other words, in this scenario there is now an independent branch of ``breather-precursor'' solutions of the more delocalized type which still exist (but are destabilized) at frequencies 
at which the actual on-site breather branch has already passed its maximum norm and reshapes towards a single-site excitation.
Interestingly, for larger $|\mu|$, the precursor branch in Fig.~\ref{fig:N6}(a) also exhibits a local norm maximum, in a fashion similar to what was observed in Figs.~\ref{fig:N2}(a) and \ref{fig:N5}(a) for the on-site breather branches.
In the vicinity of this point, its profile starts to rearrange towards a split-peak structure, see Fig.~\ref{fig:N6}(b).
The $k$-space analysis shows that this is accompanied by a pronounced admixture of quasi-momentum components near $q_2 = 4\pi/N$ as given by Eq.~(\ref{eq:qmin}),
and the squared central amplitude of the precursor branch is in the vicinity of $|A|^2_{(2)}=2 t_1/(-U) \sin^2(2\pi/N)=1.5$ as in Eq.~(\ref{eq:crampl2}) when the peak structure starts to change.

Figs.~\ref{fig:N6}(c,d) show the corresponding linearization spectra as a function of the frequency. Increasing $|\mu|$ away from the linear limit leads to
a degenerate pair of linearization modes splitting off from the phonon band and reaching zero at the critical point of maximum norm of the breather branch.
From here on, the two branches behave differently. For the on-site breather itself, Fig.~\ref{fig:N6}(c), one of the two aforementioned modes is reflected back to the real axis and one crosses to the imaginary axis,
causing exponential instability in the interval of negative slope of $\mathcal P(|\mu|)$. 
Moreover, the translational zero mode inherited from the linear limit quickly increases towards the phonon band in the vicinity of the critical 
frequency. In contrast, Fig.~\ref{fig:N6}(d) shows that for the breather precursor branch both degenerate linearization modes cross to the imaginary axis, rendering
the branch unstable once the breather has split off, although the slope of $\mathcal P(|\mu|)$ remains positive. When the breather precursor norm reaches its maximum, another linearization
eigenmode turns imaginary and comes back to the real axis only when $\mathcal P(|\mu|)$ has passed its minimum. Furthermore, 
the translational zero mode also rises near the point of maximum norm, but in this case assumes imaginary values, leading to an additional exponential instability.
It is worth mentioning here that for all branches of on-site breather solutions considered,
we have found that imaginary linearization eigenvalues are present in the intervals of negative $\mathcal P(|\mu|)$ slope (and only there).
This suggests the existence of a Vakhitov-Kolokolov-type stability criterion for this family of modes \cite{Vakhitov1973}.
The breather precursor branch of Fig.~\ref{fig:N6} illustrates that for other modes a positive slope of $\mathcal P(|\mu|)$ does not generally exclude exponential instability in our model,
a restriction that applies even in the nearest-neighbor DNLS equation \cite{Laedke1994}.

Our observations in the context of Fig.~\ref{fig:N6} further support the suggested analogy between plane-wave MI and the stability properties of the localized-mode branches, admitting the interpretation
that the precursor branch still has $k=0$ plane-wave characteristics after the actual breather has reached its maximum norm.
Beyond this first critical point, the breather precursor is linearly unstable, which may be thought of as the analogue of the $q_1$-MI of the band-edge wave.
Increasing the amplitude further, the next stage of the cascading MI sets in, and the precursor branch is 
prone to the third disjoint interval of $q$-instability experienced by the $k=0$ plane wave beyond $|A|^2_{(2)}$, cf. the similar structures in Fig.~\ref{fig:N5}(e),
which finally triggers the rearrangement towards a more localized multi-peak profile.

Summarizing, we have argued that for $U<0$ the branch of on-site breather solutions of Eq.~(\ref{eq:sdnls}) is closely linked to the $k=0$ lower band-edge wave and its MI properties.
The near-linear weakly localized part of the branch turns unstable when it reaches an amplitude in the vicinity of the second-stage MI amplitude $|A|_{(1)}^2$ of this band-edge wave.
This coincides with the maximum of the $\norm(|\mu|)$ curve and triggers a rearrangement of the quasi-momentum composition of the breather solution, leading to 
a modulation of the direct space profile on the length scale of $N$. Estimates of the critical norm and frequency can be obtained from suitable approximation schemes for the near-linear breather profile.

\subsection{Repulsive nonlinearity}
We now turn to the discussion of repulsive nonlinearity, $U>0$.
Here, we need to distinguish the cases of $N$ even and $N$ odd, respectively. 
The latter case is, in fact, already covered by the above discussion,
since for $N$ odd any solution of Eq.~(\ref{eq:sdnls}) at $\mu$, $U$ 
can be mapped to a solution at $-\mu$, $-U$  via the so-called staggering transformation $\psi_j \rightarrow (-1)^j \psi_j$, cf. \cite{Kevrekidis2009}.
The associated symmetry of the dispersion curves has already been pointed out in Fig.~\ref{fig:setup}(b).
By virtue of this mapping, at $N$ odd the characteristic $\norm(|\mu|)$ curves of the breathers in the repulsive model are the same as in the attractive model,
with the only difference of an additional relative sign between adjacent sites in the profiles.
In particular, the on-site breather branch does not start from the lower band edge at $k=0$ in the linear limit,
but instead from the upper band edge at $k=\pi$.

In contrast, the staggering transformation does not go through for $N$ even and $t_N > 0$,
since in this case it switches the relative sign between the hopping terms.
Again, a fundamental consequence of this has already been observed in Fig.~\ref{fig:setup}(b),
namely the asymmetry between the lower and the upper band edges. 
Generically (unless $t_N$ is small), there are two degenerate maxima of $\mu_0(k)$ for $N$ even,
which are located inside the Brillouin zone and thus correspond to complex plane waves.
For repulsive nonlinearities, the on-site breather branch is expected to approach the upper band edge in the linear limit,
but given the split character of this band edge a more detailed investigation is required.

\begin{figure}[ht!]
 \hspace{1mm} \includegraphics[height=2.99cm]{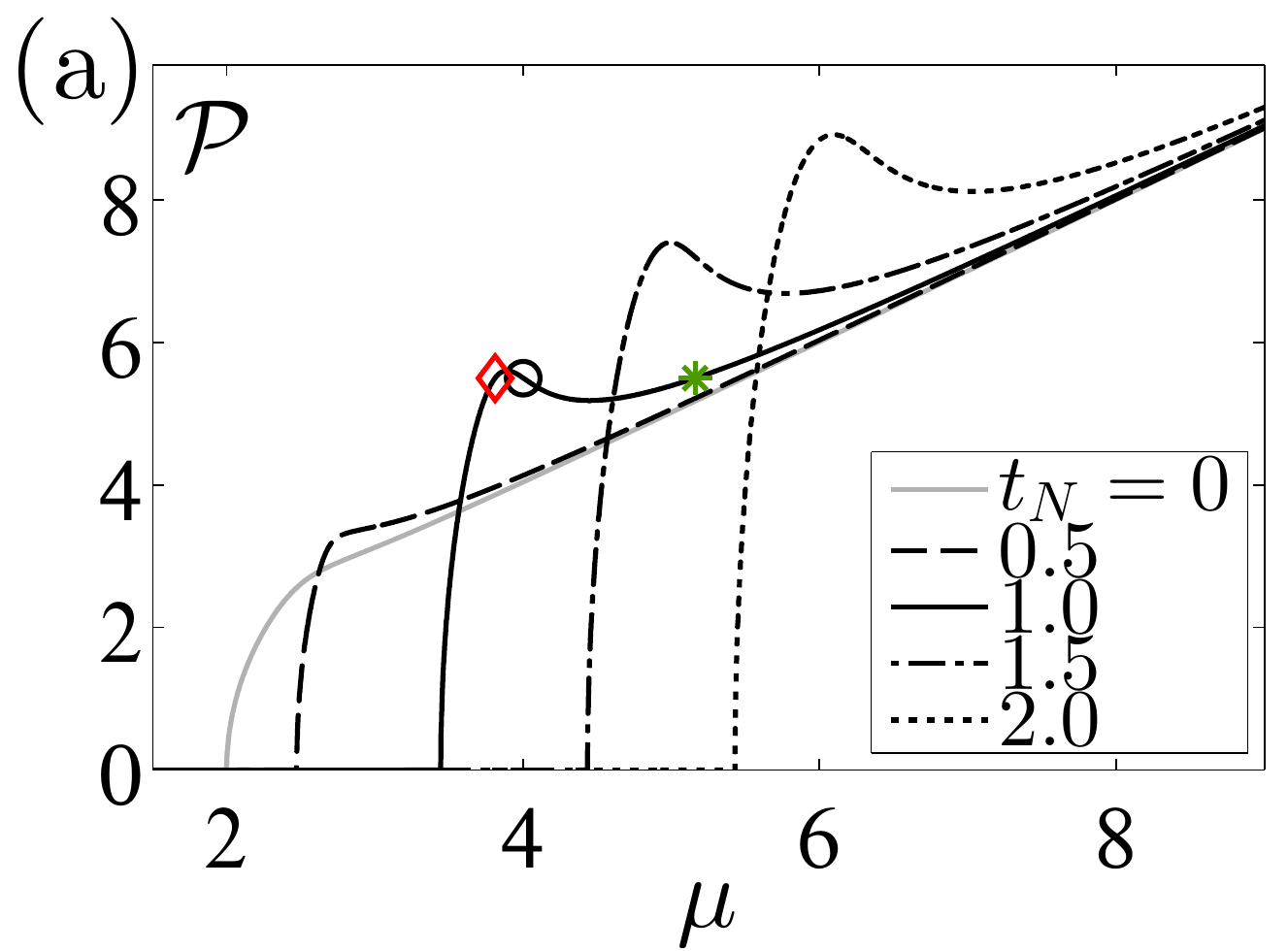}\, \includegraphics[height=2.99cm]{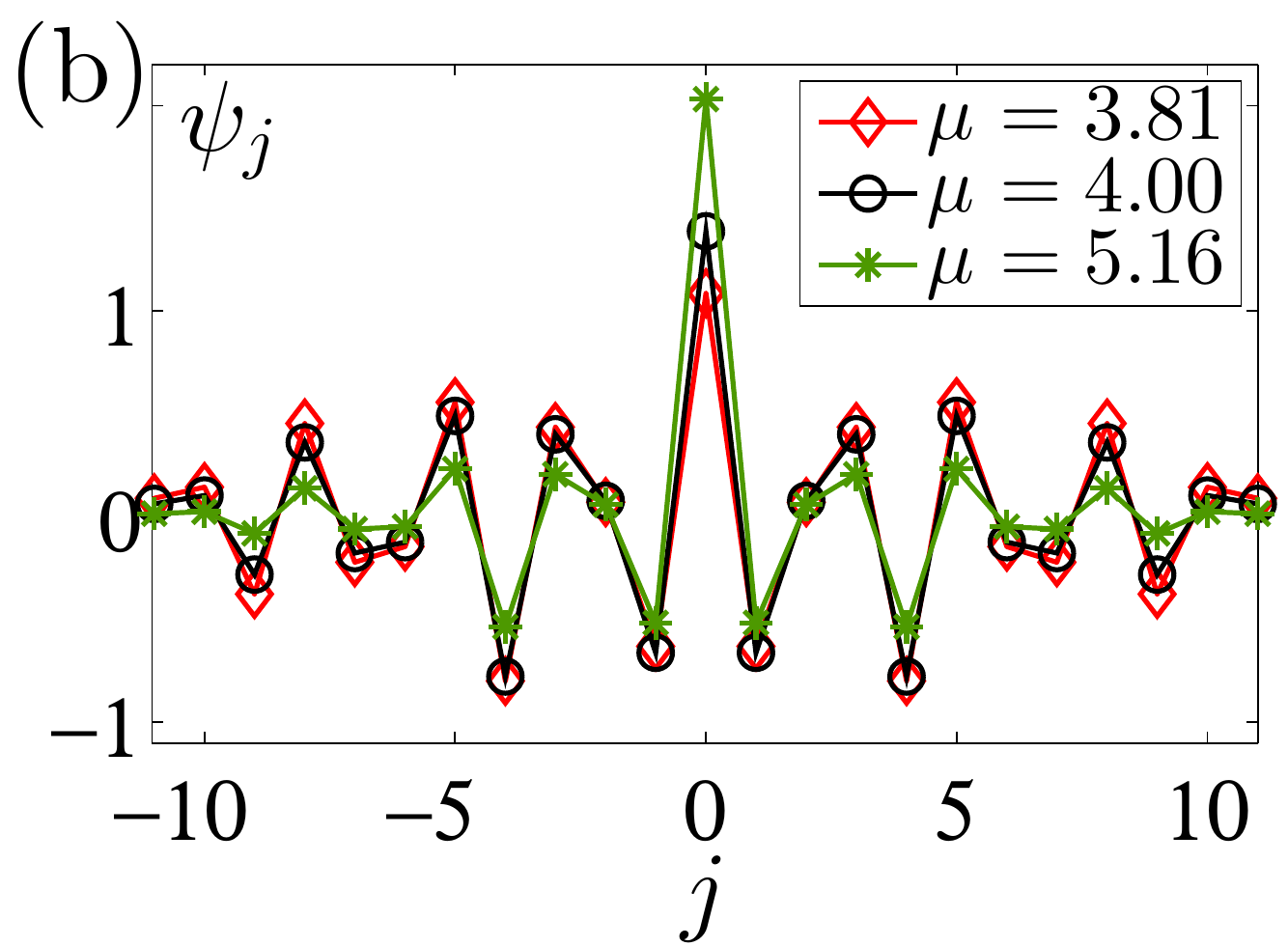}\\
\includegraphics[height=2.99cm]{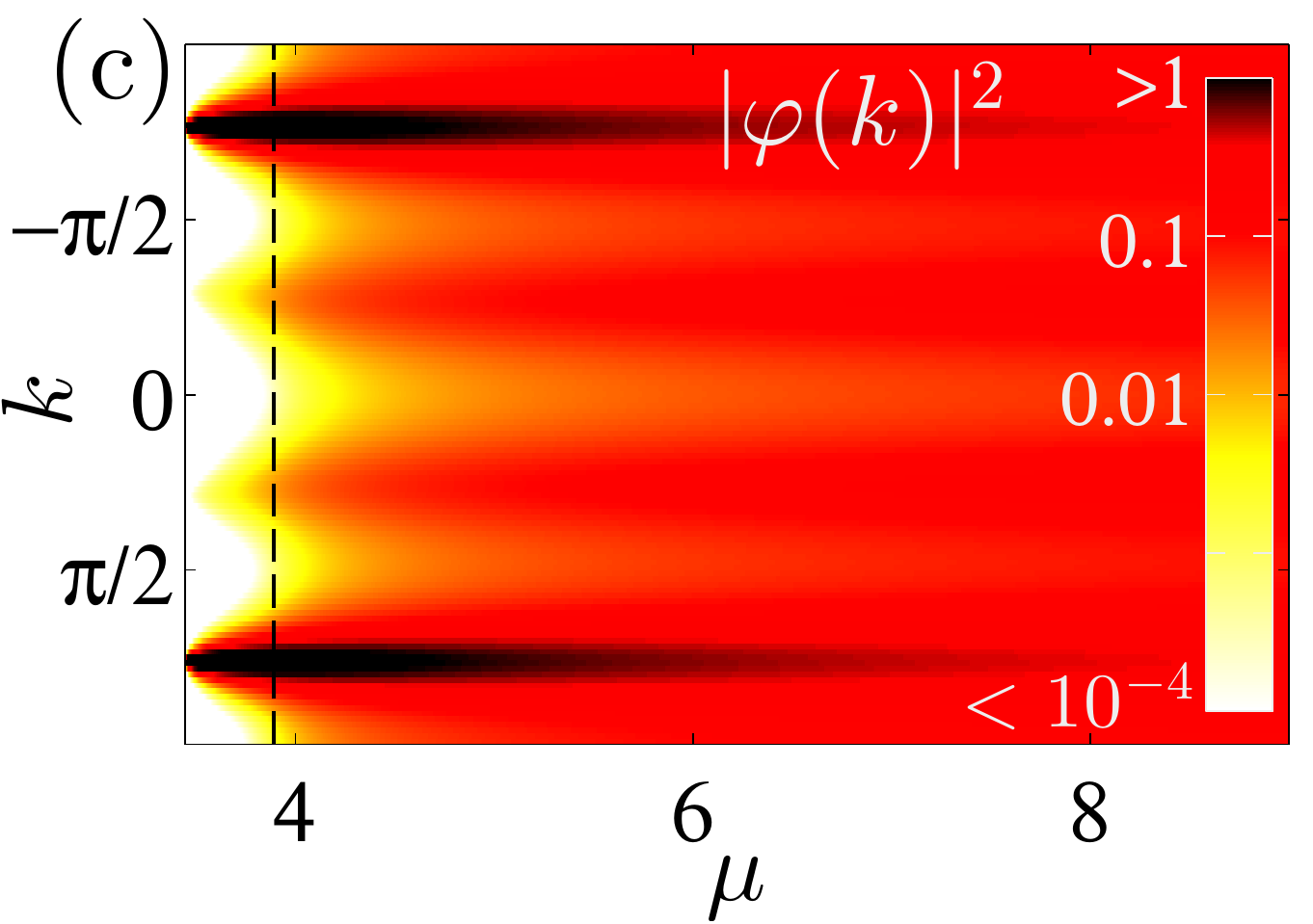} \includegraphics[height=2.99cm]{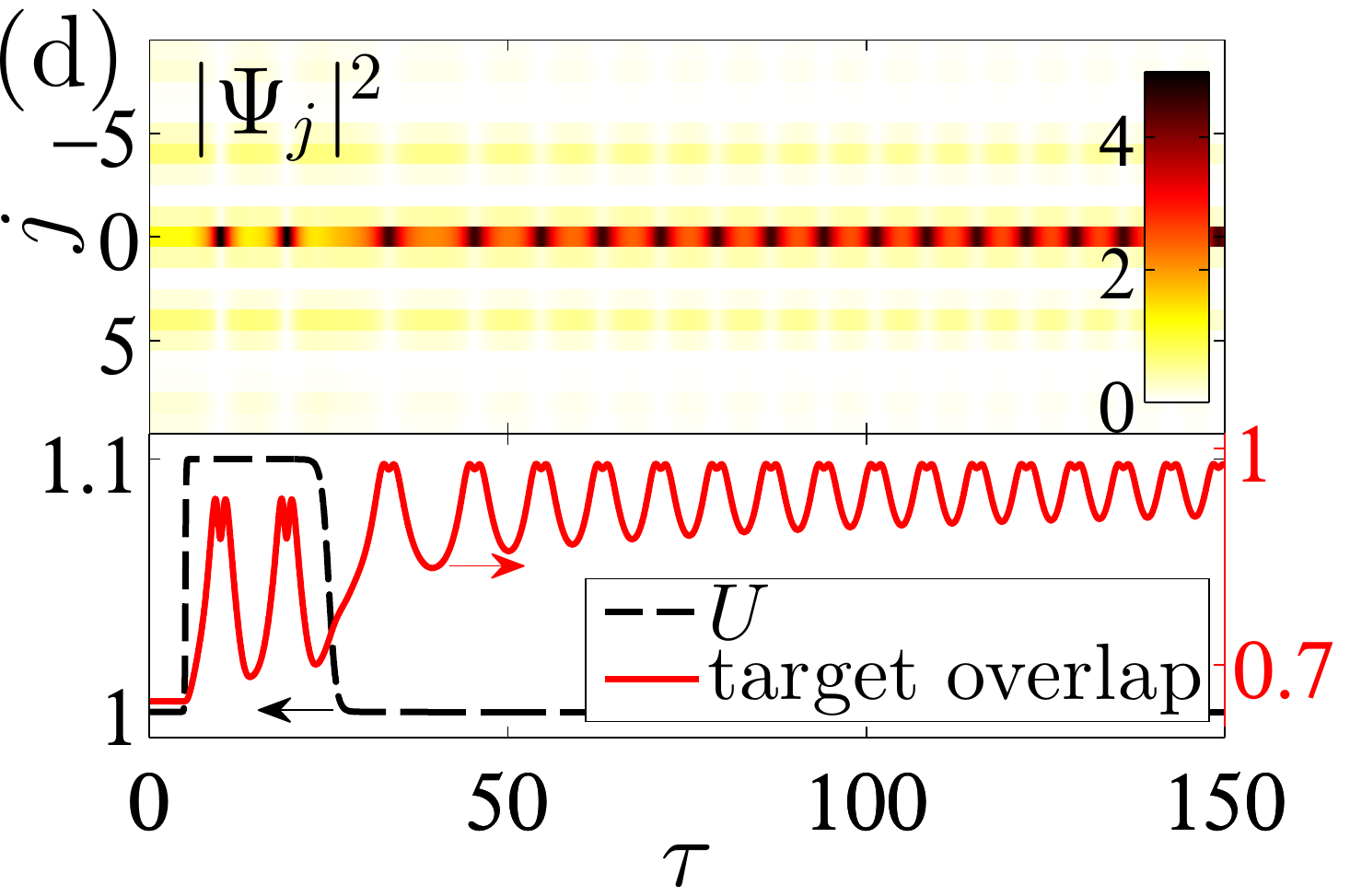} \\
\caption{\label{fig:N4}(Color online) (a) $\norm(\mu)$ curves for the on-site breather branch of solutions at different $t_N$. (b) Direct space profiles at $t_N=1$ and values of $\mu$ as indicated by the markers in (a).
(c) Normalized discrete Fourier transform of the branch of solutions at $t_N=1$ and varying $\mu$. The dashed line denotes the position at which $\norm(\mu)$ has its local maximum.
Throughout $N=4$, $t_1=1$, $U=+1$.
(d) Switching from the broad [red diamond markers in (b)] to the narrow breather [green asterisk markers in (b)] by quenching the nonlinearity coefficient $U=U(\tau)$ as shown in the bottom panel
(dashed black line, left axis). The normalized overlap with the narrow breather target state finally oscillates at above 90\% (solid red line, right axis).
}\end{figure}
Fig.~\ref{fig:N4}(a) shows a set of numerically obtained $\norm(\mu)$ curves at $N=4$ and $U=1$.
Again, these exhibit the characteristic bistability shape, 
with local extrema appearing beyond a critical $N$-th neighbor hopping ($t_N \gtrsim 0.57$ here),
and the interval of negative slope of $\norm(\mu)$ is associated with linear instability.
We find that such bistability is a generic feature also of the repulsive model at $N$ even, 
with the notable exception of $N=2$ where we see no indications of a local maximum-minimum structure of $\norm(\mu)$ even for large values of $t_N$.
In contrast to what was observed in the attractive case, the maxima of the $\norm(\mu)$ curves are broader and do not show 
the tendency of cusp formation for $U>0$ and $N$ even. 
Also, the breather profiles, as shown in Fig.~\ref{fig:N4}(b) for $t_N=1$, preserve their overall shape when crossing the norm maximum and minimum, respectively.
This is also reflected by the more smooth dependence of the Fourier coefficients on $\mu$, see Fig.~\ref{fig:N4}(c).
In particular, the maximum of $\norm(\mu)$ is not accompanied by a sudden increase of certain $k$-components in the spectrum;
there are notable peaks near $\pm \pi/4$, which however start in the immediate vicinity of the linear limit already.
Furthermore, the role of the split band edge is clearly visible in the Fourier spectrum. 
Close to the linear limit of small $\norm$ and small $\mu$, 
the on-site breather tends towards a superposition of two plane waves with wave numbers near $k=\pm 3\pi/4$,
the two degenerate maxima of the linear dispersion $\mu_0(k)$, cf. Fig.~\ref{fig:setup}(b).
This is a real superposition of the two complex band edge waves, with a maximum at the site of the breather's peak,
and the breather profile remains real throughout.
In contrast to the attractive case, where each breather branch started out as a weakly enveloped wave packet showing
no modulation on the length scale of $N$, here the nontrivial position of the split upper band edge induces such a modulation
right from the linear limit. 
Thus, in the repulsive case the overlap between the two stable solutions on the on-site breather branch having the same norm 
(one at smaller $\mu$ than the maximum of $\norm(\mu)$, one at larger $\mu$ than the minimum) is much larger than in the attractive case. 
This suggests the possibility of dynamical switching between the stable breather configurations in the presence of repulsive nonlinearity and even $N\geq 4$,
similar to what was demonstrated for attractive DNLS models with monotonically decaying long-range hopping in \cite{Johansson1998}.
Indeed, we have found such robust switching to be possible in our model, using for instance local phase kicks (requiring single-site addressability) as in \cite{Johansson1998},
but also via quenches of the global nonlinearity coefficient $U$, an example of which is shown in Fig.~\ref{fig:N4}(d).
Here, the initial localization dynamics triggered by ramping up $U$ is found to be robust against variations of the quench protocol, 
while the subsequent reduction of $U$ back to its initial value needs more fine-tuned timing in order not to fall back towards
the initial state. In line with the above discussion, we find that similar attempts of switching usually fail in the cases of $N$ odd and/or attractive nonlinearity,
where the two coexisting stable solutions exhibit qualitatively different profiles.

\begin{figure}[ht!]
 \includegraphics[height=2.99cm]{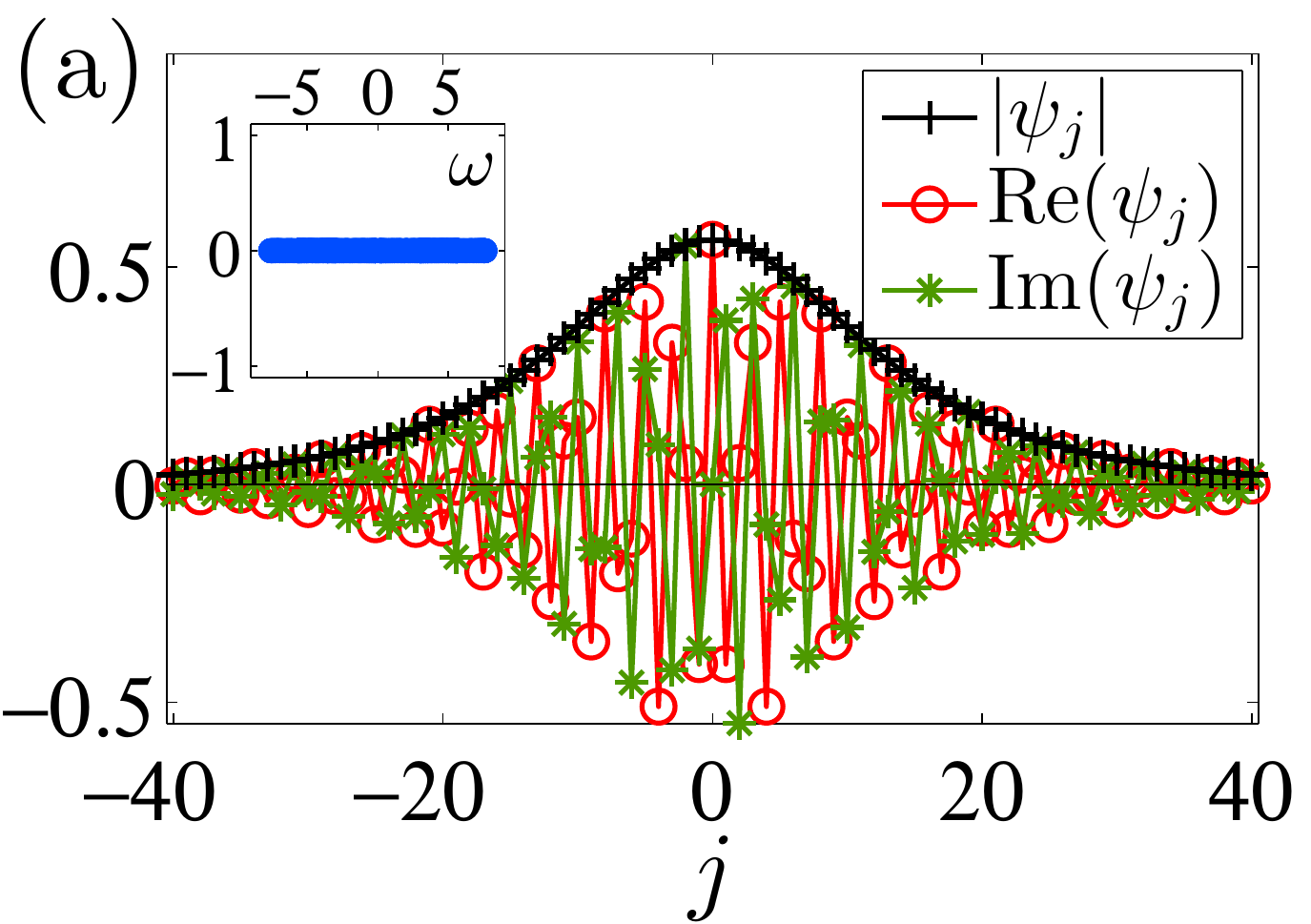} \includegraphics[height=2.99cm]{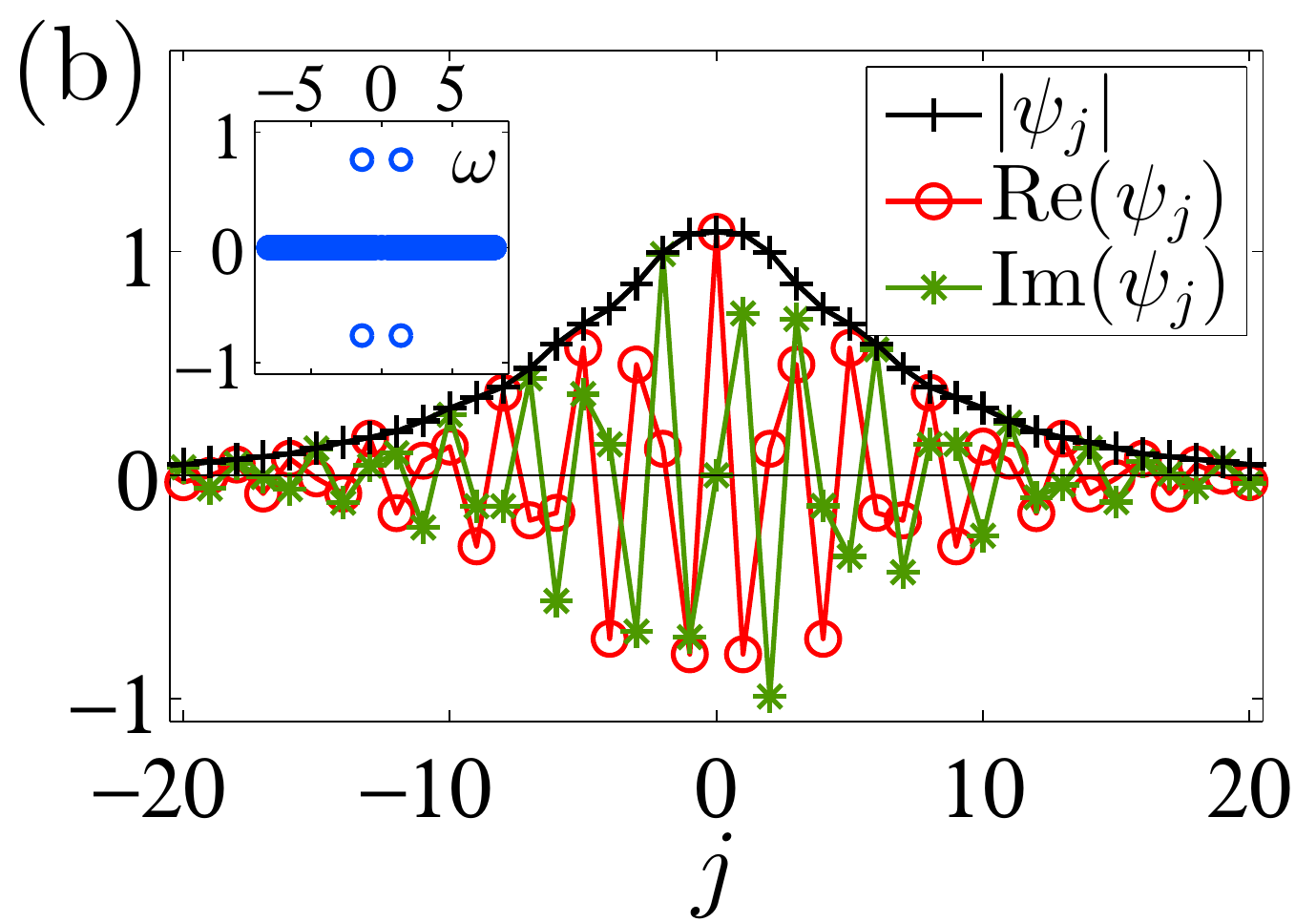}\\
\includegraphics[height=2.99cm]{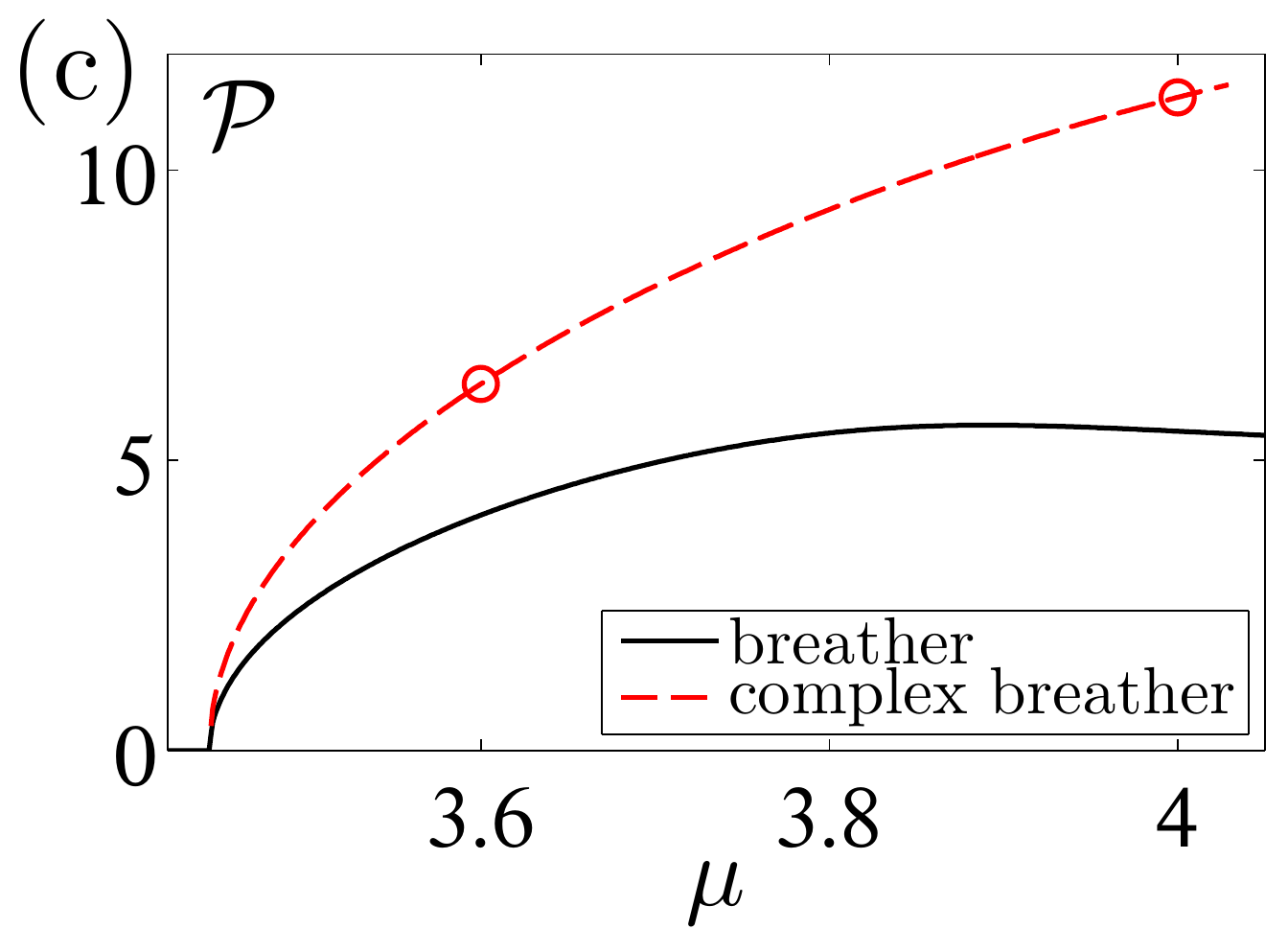} \hspace{.1mm} \includegraphics[height=2.99cm]{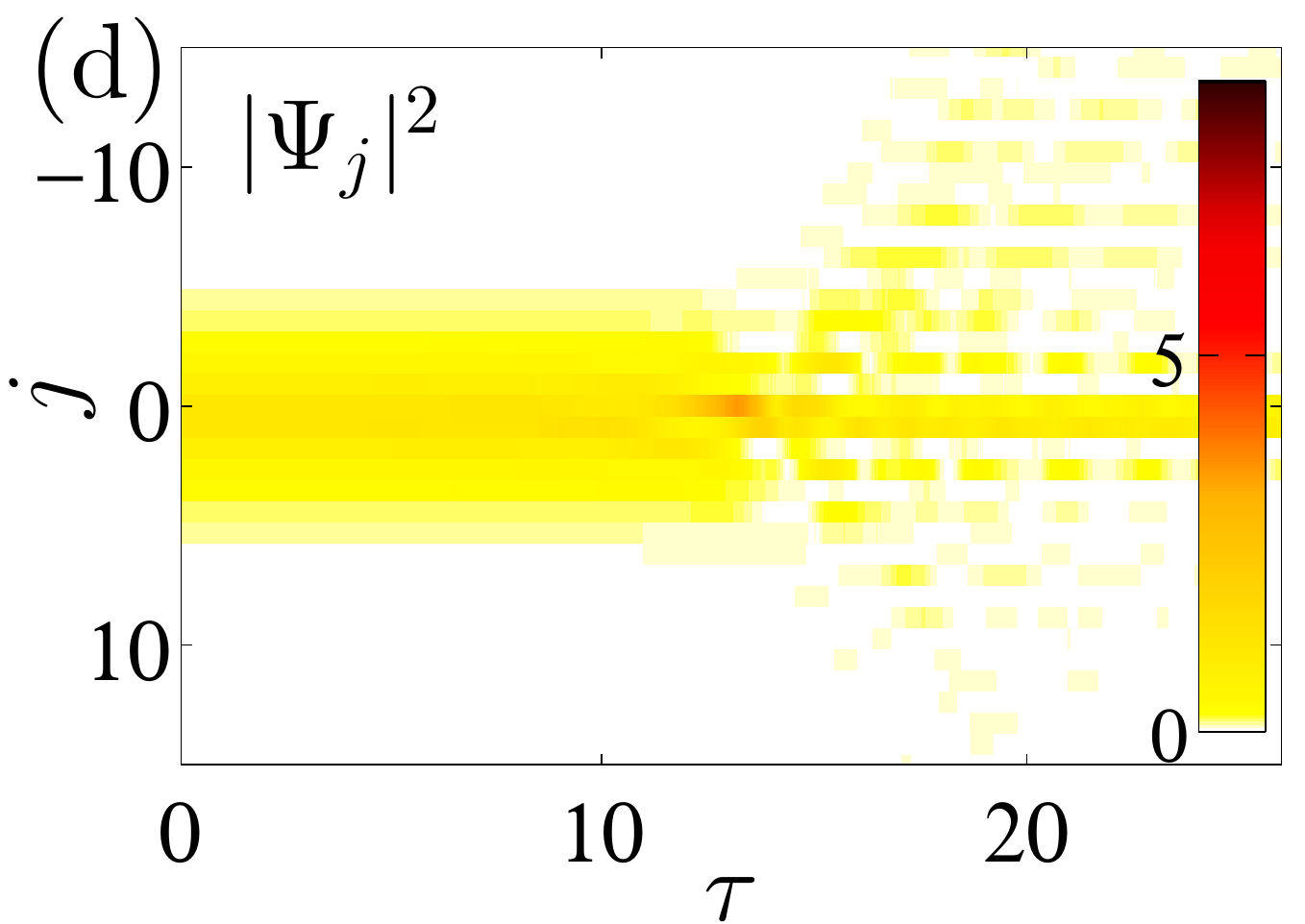}\\
\includegraphics[width=0.45\textwidth]{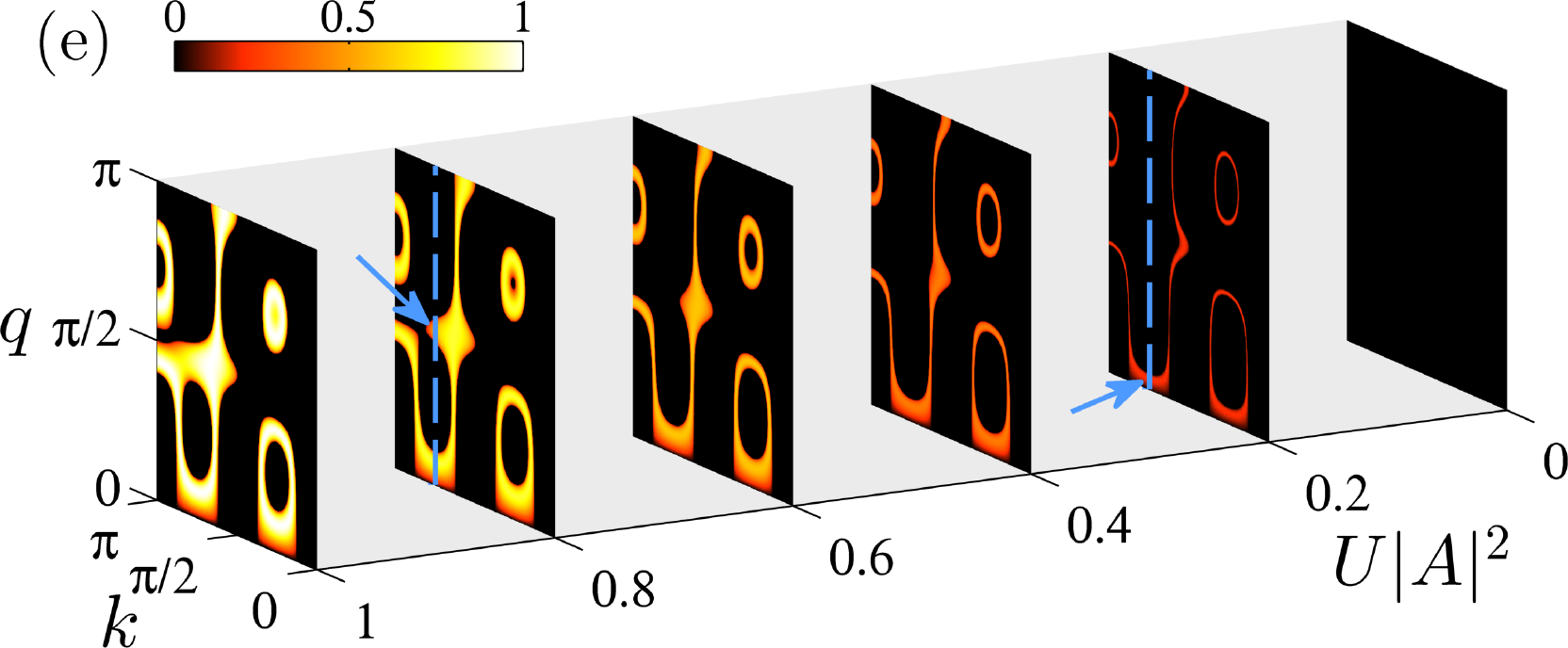}
\caption{\label{fig:comb}(Color online) Complex breather profiles for $N=4$, $t_N=1$, $L=1000$ at (a) $\mu=3.6$, (b) $\mu=4.0$. The insets show the corresponding linearization spectra in the complex plane, indicating stability and oscillatory instability, respectively. 
(c) $\norm(\mu)$ curves of the on-site breather and the complex breather branch near the linear limit at $t_N=1$. Circular markers denote the parameter values at which profiles are shown in (a,b).
(d) Decay dynamics of the mode shown in (b).
(e) MI instability diagram for $N=4$, $t_N=1$ and repulsive nonlinearity, colors encode $|\text{Im}\, \omega_\pm|$ from Eq.~(\ref{eq:om}). The dashed blue lines indicate the position of the upper band edge, which experiences a two-stage MI.
Throughout $t_1=1$, $U=+1$.}\end{figure}

The fact that for $U>0$ and $N$ even the on-site breather branch asymptotes towards a superposition of two plane waves in the linear limit indicates
that the plane-wave MI analysis as used in the attractive case is not straightforward to apply here, since it provides no stability information about such plane wave superpositions.
Still, even in the repulsive case there are other localized modes which are closer to single-$k$ plane waves and can be studied with the MI-based arguments.
To show this, let us focus on the individual complex band-edge waves in the following.
From the general discussion in Sec.~\ref{sec:MI}, the dispersion curvature at the band maximum indicates 
that the band-edge wave will be unstable towards small-$q$ MI at infinitesimal amplitudes.
Again, this is accompanied by the existence of a localized branch of solutions 
which now inherits the complex structure of the parental band-edge wave. Two profiles of such ``complex breather'' solutions
are shown in Figs.~\ref{fig:comb}(a,b). The corresponding $\norm(\mu)$ curve is provided in Fig.~\ref{fig:comb}(c).
Increasing $\mu$ away from the linear limit, the complex breather branch also shows a tendency to localize, while the norm monotonically increases.
Beyond a critical $\mu$ (given by $\mu \approx 3.88$ in this example), a substantial oscillatory instability appears, see the insets in Figs.~\ref{fig:comb}(a,b). Increasing $\mu$ further, the amplitudes at the 
three central sites of the complex breather asymptote to the same value and eventually the numerical continuation fails, see Fig.~\ref{fig:comb}(c), suggesting 
a bifurcation scenario in which the branch ceases to exist.
Fig.~\ref{fig:comb}(d) provides an example of the dynamics triggered by perturbing the complex breather in its unstable frequency interval, 
showing decay towards a highly localized single-site configuration.

Now to understand the oscillatory destabilization of the complex breather branch, we can again invoke the plane-wave MI analogy.
Fig.~\ref{fig:comb}(e) shows the corresponding MI diagram of the repulsive model at $N=4$. 
It can be seen that the band-edge wave at $k\approx 3\pi/4$ indeed experiences a two-stage MI, 
with the aforementioned small-$q$ instability at infinitesimal amplitudes and a detached range of MI emerging near $q=\pi/2$ below $U|A|^2=0.8$.
For comparison, when the oscillatory instability of the complex breather branch emerges, its central amplitude $|\psi_0|^2 \approx 0.9$.
Furthermore, while Fig.~\ref{fig:comb}(e) only shows the imaginary part of the MI frequency, Eq.~(\ref{eq:om}) also predicts a non-zero real part of the unstable frequencies of around $\pm 1.4$ now,
in good agreement with the real part of the observed oscillatory instability eigenvalues, see the inset of Fig.~\ref{fig:comb}(b).

Summarizing, for odd values of $N$ the staggering transformation maps the repulsive to the attractive model and the discussion of the previous subsection still applies.
For $N$ even, the MI analogy cannot be employed directly to predict the stability properties of the fundamental on-site breather due to the split upper band edge,
but it is still of use in analyzing other localized solutions, particularly the weakly localized breather-type states that stem from the complex band-edge waves and feature nontrivial phase profiles.

\section{Dimensional crossover}
\label{sec:2D}
In this section we explore a different facet of the helicoidal DNLS Eq.~(\ref{eq:dnls}), namely the limit of large $N$.
Geometrically, this means that the number of sites per winding is increased while simultaneously the pitch-to-radius ratio is decreased such 
that $t_1$ and $t_N$ remain of the same order.
While this is not immediately obvious from the equation itself,
thinking of the associated helix geometry as shown in Fig.~\ref{fig:setup}(a)
suggests that in this limit the lattice will locally approach a two-dimensional (2D) square lattice, see also \cite{Xiong1992}.
Regarding the on-site breather solutions, this may seem surprising,
since it is well-known that in a 2D square lattice these breathers have properties very different from their 1D counterparts.
In particular, in the 2D-DNLS there is a finite threshold norm below which no breathers exist \cite{Flach1997,Weinstein1999}, see also \cite{Cuevas2009}.
Approaching the linear limit from large frequencies (with a correspondingly highly localized solution), the 2D $\norm(|\mu|)$ curve exhibits a global minimum and then rises again,
asymptoting towards a finite value when the frequency reaches the band edge.
Here, for $U<0$ the solution delocalizes and eventually reduces to the so-called Townes soliton \cite{Chiao1964}, the unique bright soliton solution of the 2D attractive nonlinear Schrödinger equation in the continuum.
Furthermore, the 2D model is again symmetric under a staggering transformation, such that the $\norm(|\mu|)$ curves are invariant under $U \rightarrow -U$.
It is interesting to see if and how these features emerge from the 1D-DNLS model with helicoidal hopping when $N \rightarrow \infty$.

\begin{figure}[ht!]
 \includegraphics[width=0.45\textwidth]{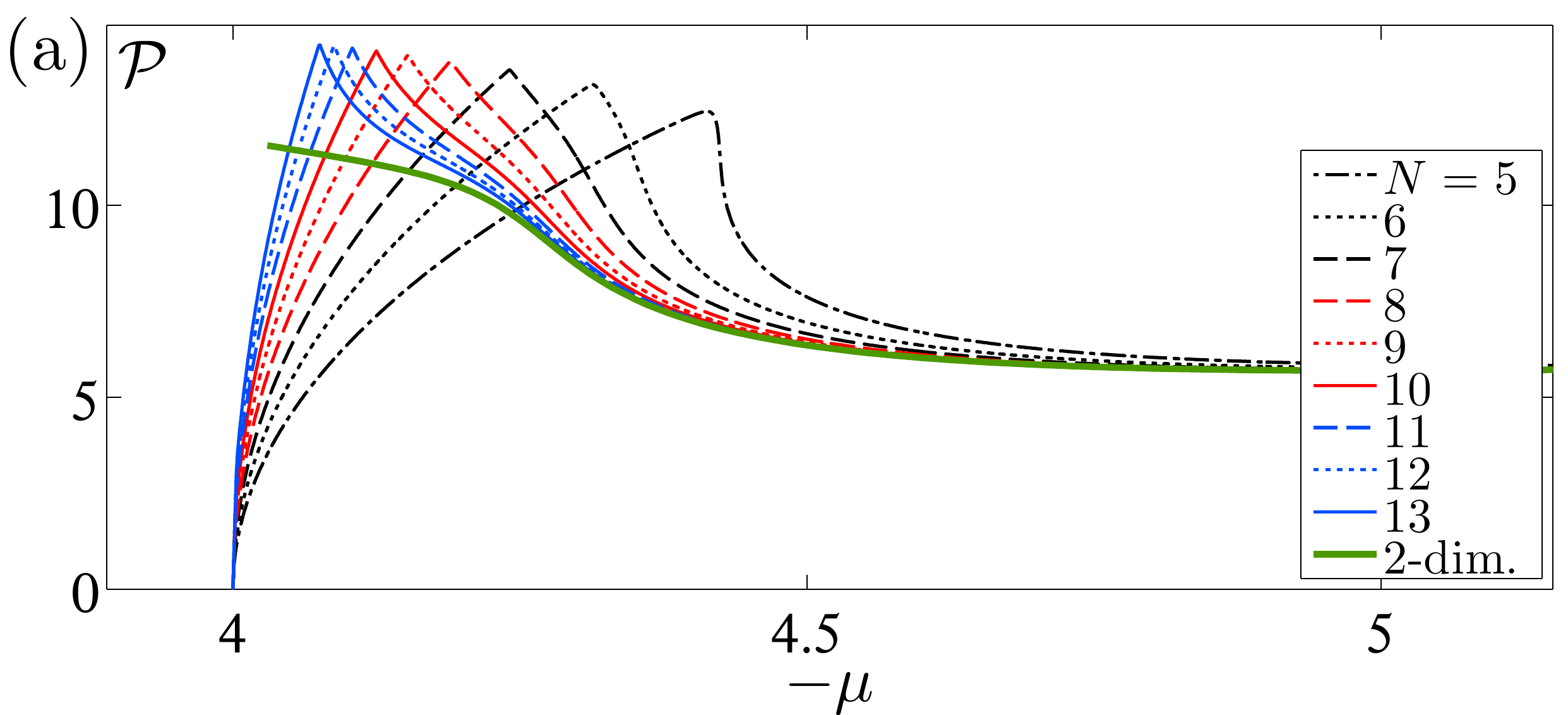} \includegraphics[width=0.45\textwidth]{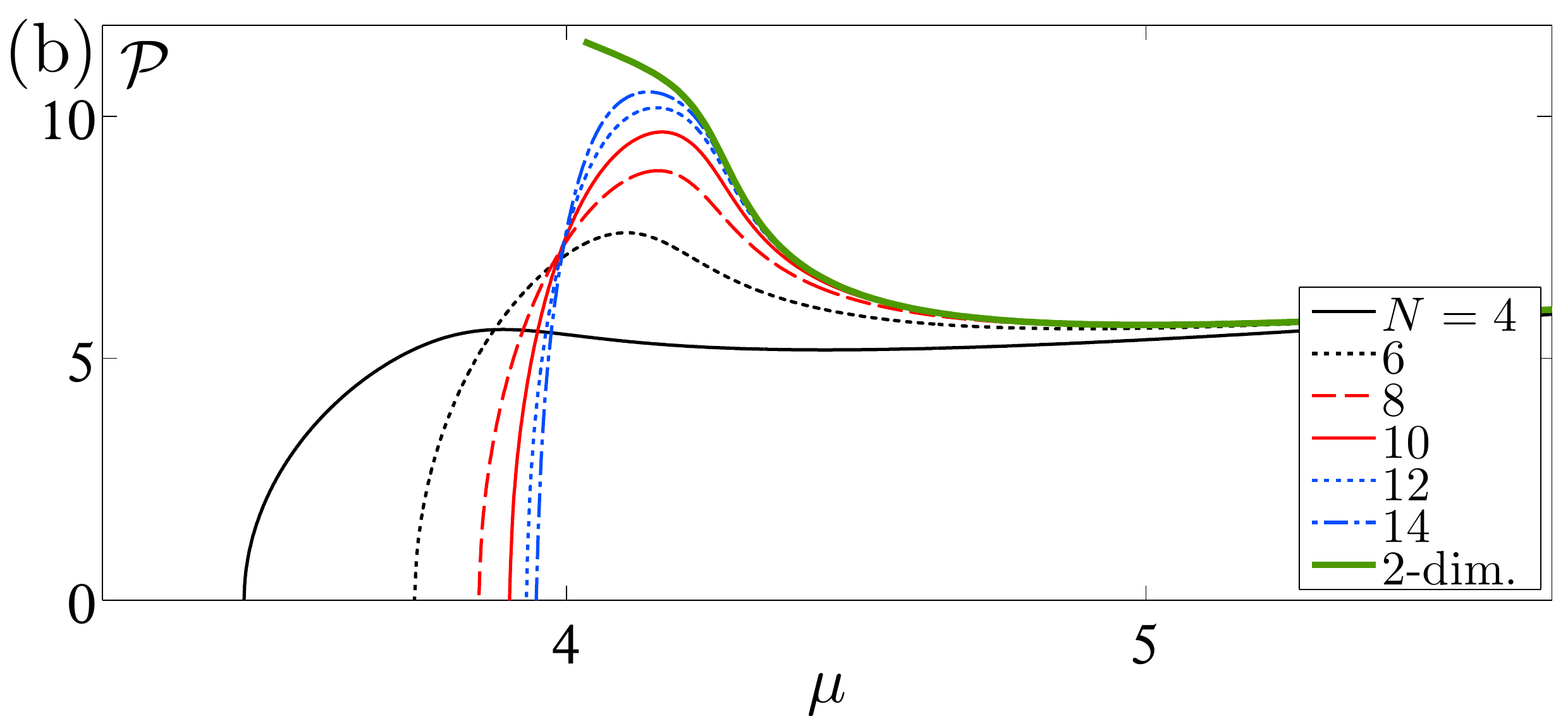}
\caption{\label{fig:2D}(Color online) Effective dimensional crossover in the on-site breather $\norm(|\mu|)$ curves at large $N$. 
(a) Attractive nonlinearity, $U=-1$. (b) Repulsive nonlinearity, $U=+1$.
The $\norm(|\mu|)$ curve obtained for the on-site breather branch in the 2D-DNLS equation for $|U|=1$ (independently of the sign) is shown as a solid green line. 
Throughout $t_1=t_N=1$.}\end{figure}
In Fig.~\ref{fig:2D}, we present numerical results showing how the $\norm(\mu)$ curves of the breather solutions of Eq.~(\ref{eq:sdnls}) asymptote to the 2D result as $N$ is increased.
Expectably, the norm curves coincide best at large frequencies, where the breather solution is well localized and only
senses the immediate vicinity of its peak site, which already for moderate values of $N$ has the square lattice structure.
A value of $N \approx 8$ is sufficient to fully match the minimum of the 2D $\norm(|\mu|)$ curve on the scale of Fig.~\ref{fig:2D}.
When moving closer to the linear limit, the breathers tend to delocalize and eventually are affected by the helix periodicity that deviates from the square lattice.
For $U<0$ (and for $U>0$ and $N$ odd), in this region near the linear limit the solutions of the helicoidal DNLS equation have significantly larger norms than their 2D analogues,
before the $\norm(\mu)$ curve sharply drops to zero when $\mu$ approaches the band edge.
In contrast, for $U>0$ and $N$ even, the helicoidal DNLS tends to underestimate the norm of the 2D solution.

Even for large $N$, the $\norm(|\mu|)$ curves in the helicoidal DNLS exhibit no threshold, but still tend towards zero norm in the linear limit.
Yet, the critical value of $|\mu|$ at which the norm reaches its maximum approaches the band edge with increasing $N$, 
such that the frequency interval in which breathers with small norms exist becomes increasingly narrow, accounting for the emergence of an effective threshold.
Remarkably, this is correctly predicted by the MI analysis of Sec.~\ref{sec:br}.
Both Eq.~(\ref{eq:app1a}) and Eq.~(\ref{eq:app2a}) predict that $\mu_{\rm cr} \rightarrow -2(t_1+t_N) = \mu_0(k=0)$ as $N \rightarrow \infty$.
Even more, Eq.~(\ref{eq:app1b}) and Eq.~(\ref{eq:app2b}) predict that the critical norm approaches the finite value $\norm_{\rm cr} \rightarrow 4\pi\sqrt{t_1 t_N}/(-U)$.
With $t_1=t_N=-U=1$, this results in $\norm_{\rm cr} = 4\pi \approx 12.6$, remarkably close to the actual norm of the Townes soliton, $\norm_{\rm Townes} \approx 11.7$ \cite{Weinstein1983},
the bright soliton solution of the 2D attractive nonlinear Schrödinger equation that the 2D-DNLS equation asymptotes to near the linear limit.

\section{Conclusions}
\label{sec:con}
Motivated by lattice systems with an underlying helix geometry,
we have investigated a quasi-1D discrete nonlinear Schrödinger model with hopping terms to the first and a selected (potentially remote) neighbor at index distance $N$.
The modulational instability analysis revealed peculiar cascades of plane-wave destabilization with increasing nonlinearity.
We argued and presented numerical indications that these modulational instability features can be linked to the stability properties of near-linear localized modes in this model.
In particular, we suggested that for attractive nonlinearities the maximum in the norm-vs.-frequency curve of the fundamental on-site breather branch of solutions
can be viewed as a remnant of the multi-stage modulational instability of the plane wave at the lower band edge.
Adopting this point of view provided simple estimates for the critical parameters where the breather branch destabilizes; 
these were seen to be in qualitative (and, within the expected limited accuracy, also quantitative) agreement with the numerical findings.
A discussion of certain complex localized solutions in the presence of repulsive nonlinearity demonstrated the versatility 
and generality of the modulational-type instability analysis of such modes.
Finally, we investigated the large-$N$ limit of the model, where, intuitively, it locally approaches the 2D discrete nonlinear Schrödinger equation on a square lattice.
Clear indications of this type of dimensional crossover could be observed in the properties of the localized breather solutions.
In particular, it was seen how the breather bistability at small $N$ transforms into the threshold behavior expected in the effective-2D limit of large $N$.
This is correctly predicted by the modulational-instability based arguments, which even provide a good approximation to the norm of the Townes soliton in this limit.

While in this work we have restricted ourselves to commensurate helix geometries, accounting only for a single inter-winding hopping term,
it can be expected that the methods outlined here are relevant to a wider class of discrete nonlinear Schrödinger models with isolated long-range hopping terms.
A particularly interesting direction for future studies is an extension to multi-strand helix lattices, which arguably are more accessible for implementations in ultracold-atom experiments than the 
single-strand helix \cite{Reitz2012,Stockhofe2015}. 
Furthermore, given the emergence of the helicoidal discrete nonlinear Schrödinger equation as an envelope approximation of more complex nonlinear lattice systems,
for instance of the extended Peyrard-Bishop model of DNA \cite{Tabi2008}, 
a modulational-instability based analysis may also provide valuable information on the properties of localized solutions in such models.

\section*{Acknowledgment}
The authors thank P. G. Kevrekidis for insightful discussions. J. S. gratefully acknowledges support from the Studienstiftung des deutschen Volkes.

\appendix
\renewcommand*{\thesection}{\Alph{section}}
\section{Variational and continuum estimates}
\label{app}
It is demonstrated in the main text that in many cases the analogy to the plane-wave MI
admits an estimate of the critical central amplitude at which 
the localized breather branches of the helicoidal DNLS equation change their stability properties.
Since peak amplitude, norm and frequency of a breather branch are linked to each other,
in principle this information can be used to find the critical $\mu$ and $\norm$.
As, however, the dependence of $\mu$ and $\norm$ on the peak amplitude is usually not known explicitly,
one has to resort to approximate methods to obtain analytical expressions for the critical parameters.
In this appendix, we give the details of the derivation of Eqs.~(\ref{eq:app1a})-(\ref{eq:app2b}),
providing approximate relations between $|\psi_0|^2$, $\mu$ and $\norm$ for the near-linear breathers as 
encountered in the attractive model in Sec.~\ref{sec:br}, localized in the vicinity of $k=0$ in quasi-momentum space.

A first class of such approximations relies on using a tractable variational ansatz for the breather profile.
The stationary DNLS Eq.~(\ref{eq:sdnls}) can be obtained from varying the energy
\begin{align}
 \mathcal E [\{\psi_j\}] &= \sum_j \left (-\mu|\psi_j|^2 +\frac{U}{2}|\psi_j|^4  \right. \nonumber\\
  \hspace{-15mm}&\left. \phantom{\frac{U}{2}} -t_1 \psi_j^*(\psi_{j+1}+\psi_{j-1}) -t_N \psi_j^*(\psi_{j+N}+\psi_{j-N}) \right) \label{eq:E}.
\end{align}
Now the variational approximation consists in replacing the breather profile by an ansatz with a finite number of parameters,
such that $\mathcal E$ effectively becomes a function in few dimensions whose extrema can be determined more easily.
For the near-linear breather wave packets localized near $k=0$ in Fourier space, a commonly employed tractable ansatz is given by the exponential
$ \psi_j = B \exp(-a|j|)$,
where $a$, $B$ are real variational parameters \cite{Malomed1996,Cuevas2009a,Kevrekidis2009}.
In the NN-DNLS model, such an ansatz gives a reasonable approximation to the breather profile for arbitrary frequencies,
whereas in our model it will only apply for the delocalized part of the breather branches at $U<0$.
We insert the exponential ansatz into Eq.~(\ref{eq:E})
and eliminate the amplitude $B$ for the norm $\norm$ via
\begin{equation}
 \norm = \sum_j |\psi_j|^2 = B^2 \coth a
\label{eq:varsum}
\end{equation}
to obtain
\begin{align}
 \mathcal E(a,\norm) &= \frac{2 \norm}{\cosh a} \left(-t_1 -t_N e^{a(1-N)} \right) - \mu \norm \nonumber \\
&- \norm 2 t_N (N-1) e^{-aN} \tanh a + \norm^2 \frac{U}{2} \frac{\coth 2a}{\coth^2 a}.
\end{align}
Now the variational equations $\frac{\partial \mathcal E}{\partial \norm} = 0$, $\frac{\partial \mathcal E}{\partial a} = 0$ yield
\begin{equation}
 \mu = -2 t_1 \sech a - 2 t_N e^{-a N} (1 + N \tanh a) + \norm U \frac{\coth 2a}{\coth^2 a},
\label{eq:mua}
\end{equation}
and
\begin{eqnarray}
 \norm = &&-\frac{4 \cosh a \sinh^2 2a}{U(\sinh 4a - \sinh 2a)} \nonumber\\
 &&\quad \times \left[ t_1 + t_N N e^{-aN}(N+\tanh a) \cosh a  \right].
\label{eq:norma}
\end{eqnarray}

Eqs.~(\ref{eq:mua}) and (\ref{eq:norma}) can be combined for an equation $\mu = \mu(a)$.
This equation can then be inverted (numerically) for $a(\mu)$,
from which then the approximate $\norm(\mu)$ and $B(\mu)$ curves can be obtained.
Aiming for explicit analytical expressions, we can go one step further by noting that the breathers are broad and the decay parameter $a$ is small 
in the parameter region we are interested in. Inserting Eq.~(\ref{eq:norma}) into Eq.~(\ref{eq:mua})
and expanding to second order in $a$ yields an equation that can be analytically solved for
\begin{eqnarray}
 a(\mu) &\approx& \sqrt{\frac{-\mu - 2(t_1+t_N)}{3(t_1+N^2 t_N)} }
\label{eq:vara}
\end{eqnarray}
Then, inserting back into Eqs.~(\ref{eq:norma}) and (\ref{eq:varsum}), expanded to second order in $a$, results in
\begin{eqnarray}
 \norm (\mu) &\approx& \frac{8}{-\sqrt{3} U} \sqrt{(t_1+N^2 t_N) [-\mu - 2(t_1+t_N)]} \nonumber \\
&&+\frac{8N(N^2-1)t_N}{3U(t_1+N^2 t_N)} [-\mu - 2(t_1+t_N)], \label{eq:varnorm}\\
 B(\mu)^2 &\approx & \frac{8}{-3 U }[-\mu - 2(t_1+t_N)] \label{eq:varB}.
\end{eqnarray}
\begin{figure}[ht!]
 \includegraphics[height=2.93cm]{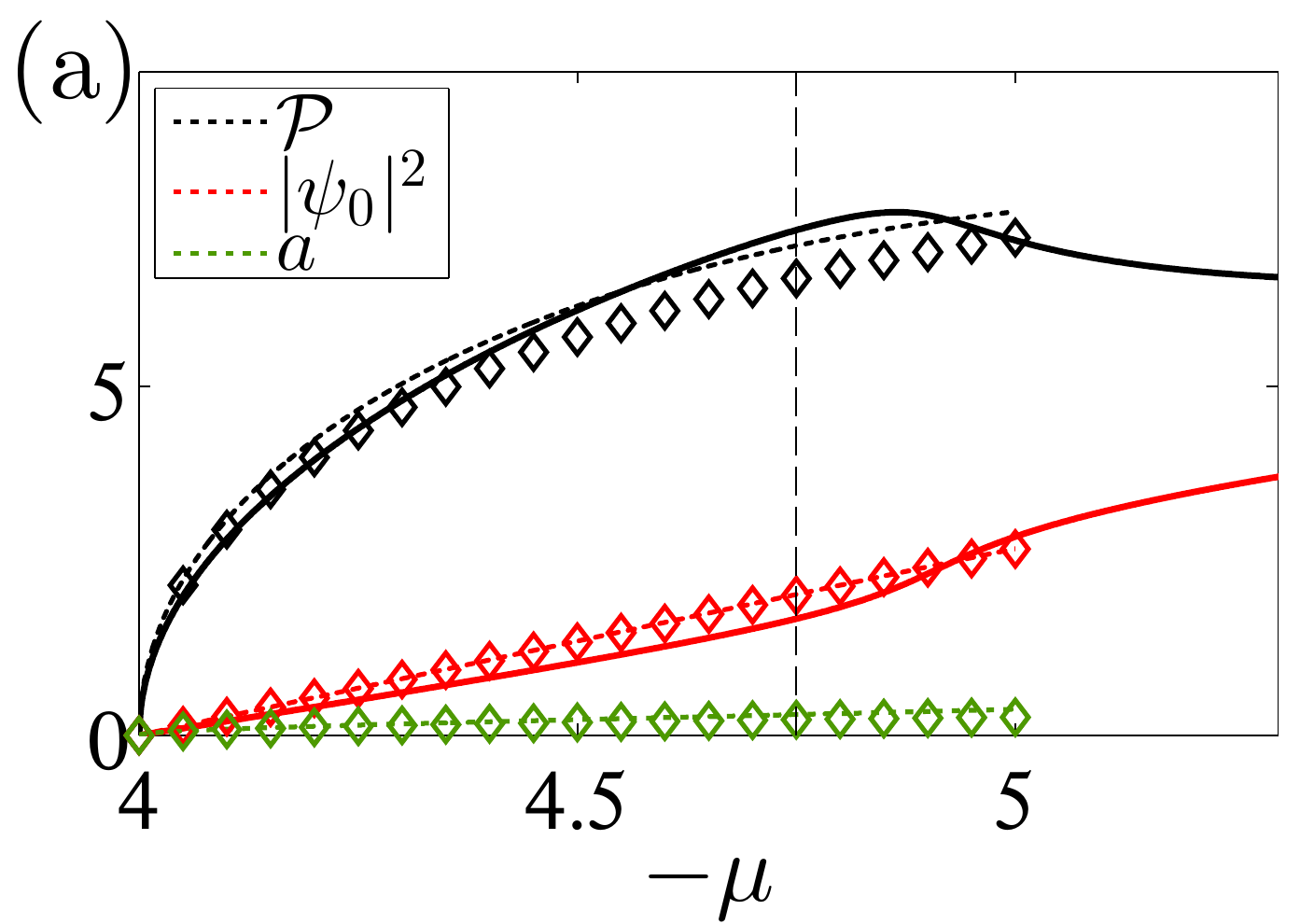}
 \includegraphics[height=2.93cm]{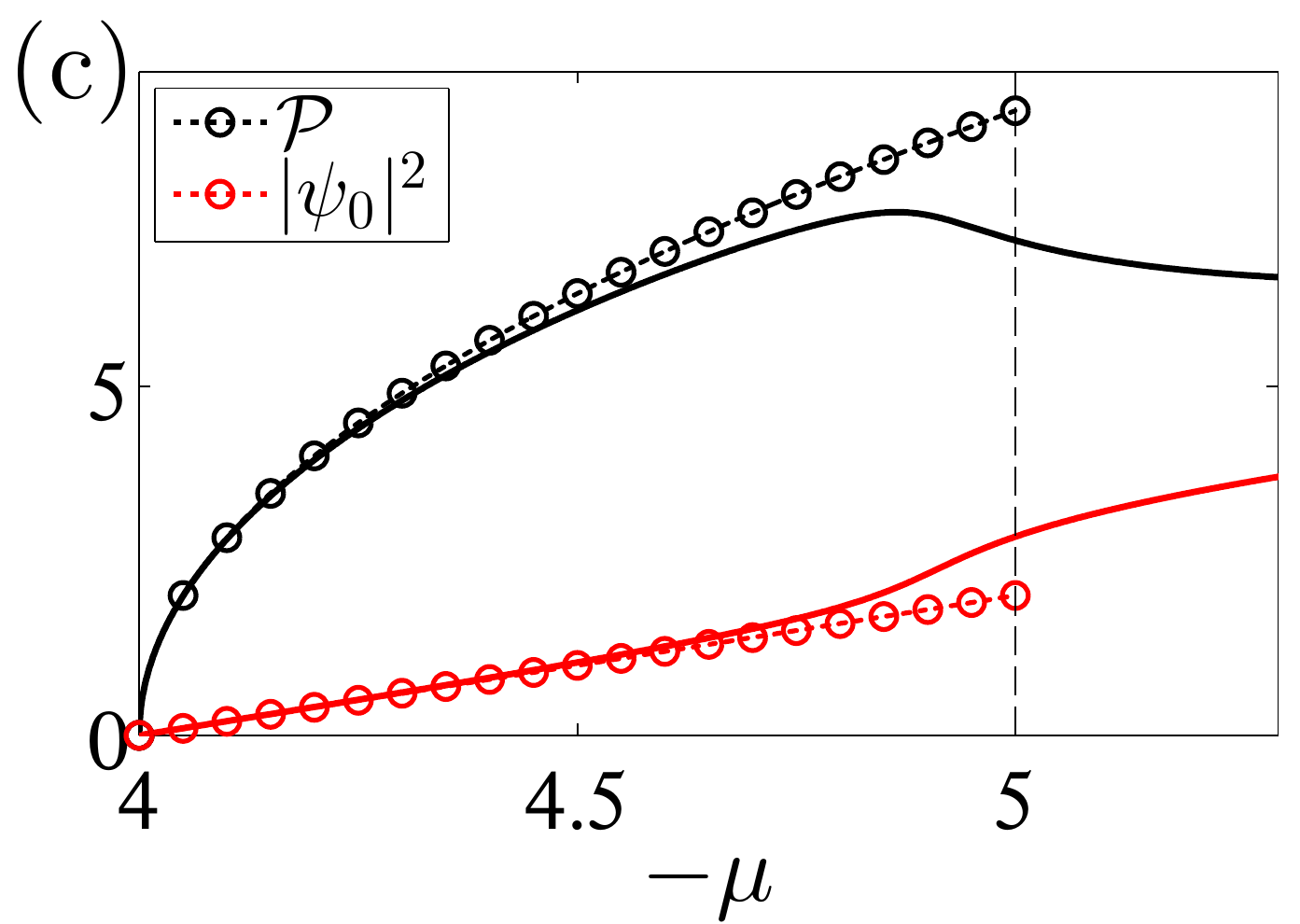}
 \includegraphics[height=2.8cm]{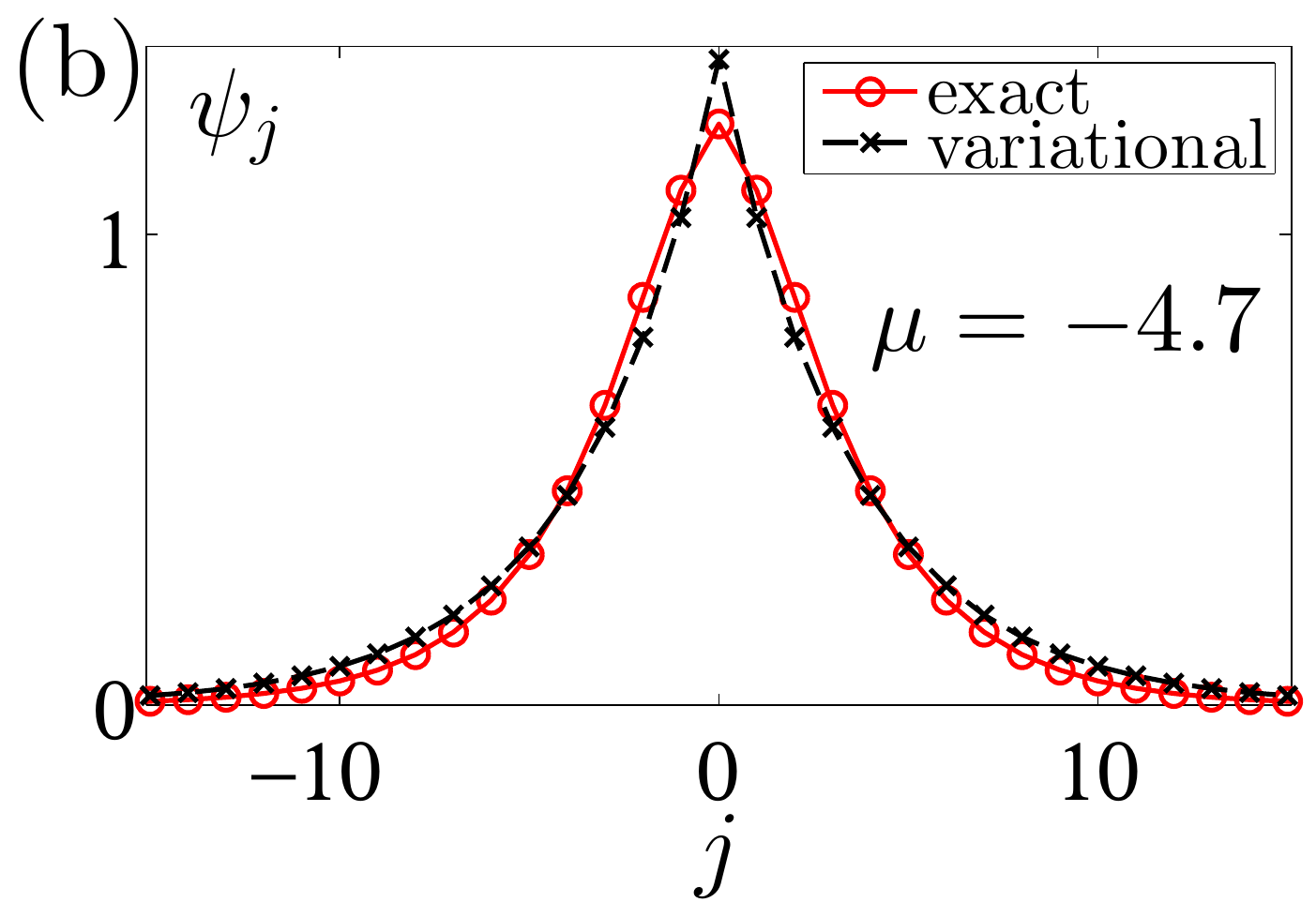}
 \includegraphics[height=2.8cm]{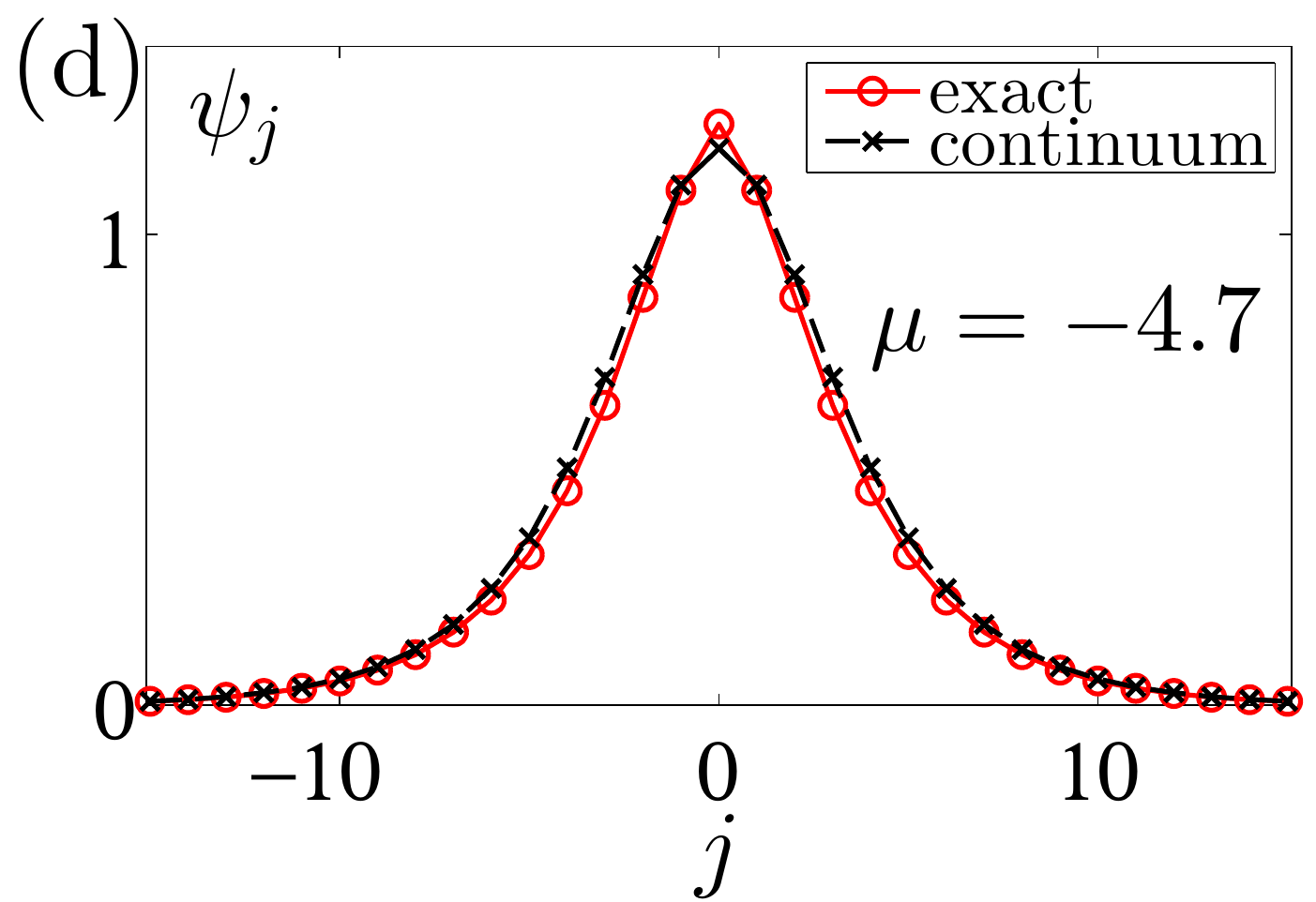}
\caption{(Color online) 
(a) Frequency dependence of the norms and peak amplitudes of the on-site breather branch:
full numerical results (solid lines), solutions to the full variational equations (\ref{eq:mua},\ref{eq:norma}) (dashed lines)
and variational results within the small-$a$ approximation Eqs.~(\ref{eq:vara},\ref{eq:varnorm},\ref{eq:varB}) (markers). 
The vertical dashed line indicates $\mu_{\rm cr}$ from Eq.~(\ref{eq:app2a}).
(b) Selected numerically exact breather profile compared to the prediction from the full variational approximation.
(c) Frequency dependence of the norms and peak amplitudes of the on-site breather branch:
full numerical results (solid lines) and expressions from the continuum approximation (dashed lines with markers).
The vertical dashed line denotes $\mu_{\rm cr}$ from Eq.~(\ref{eq:app1a}).
(d) Selected numerically exact breather profile compared to the prediction from the continuum approximation. Throughout $t_1=t_N=-U=1$, $N=2$.\label{fig:var}}\end{figure}

Fig.~\ref{fig:var}(a) compares these variational results to the numerical results for the breather branch at $N=2$ and $U=-1$,
showing relatively good agreement given the simple approximation scheme.
The additional error introduced by the small-$a$ expansion is small for the peak amplitude,
but notably larger for the norm.
Clearly, the exponential ansatz captures the overall shape of the breather solution, but tends to overestimate the peak amplitude, as can be seen from the profiles in Fig.~\ref{fig:var}(b).
Now by the reasoning of Sec.~\ref{sec:br}, MI-type instability of the breather branch is expected when its peak amplitude crosses
$B(\mu_{\rm cr})^2 = 2t_1\sin(\pi/N)^2/(-U)$, see Eq.~(\ref{eq:crampl2}), from which Eq.~(\ref{eq:varB}) predicts a critical frequency of approximately
$ \mu_{\rm cr} \approx -2(t_1+t_N) -\frac{3}{4}t_1 \sin^2(\pi/N)$ as in Eq.~(\ref{eq:app2a}) of the text. 
Inserting this back into Eq.~(\ref{eq:varnorm}) results in Eq.~(\ref{eq:app2b}).

An alternative route towards modeling the breather profile in the vicinity of the linear limit,
where it is highly delocalized and slowly varying in space, is
the continuum approximation.
Here, the discrete set of $\psi_j$ is replaced by a continuous field $\psi(x)$, and the DNLS Eq.~(\ref{eq:sdnls})
turns into an approximate nonlinear Schrödinger equation for this field,
\begin{equation}
 \tilde \mu \psi(x) = \left[ -\frac{1}{2M} \partial_x^2 + U|\psi(x)|^2 \right] \psi(x),
\label{eq:NLS}
\end{equation}
with the effective mass $1/M = \mu_0''(k)|_{k=0} = 2 (t_1+N^2 t_N)$ and $\tilde \mu = \mu + 2(t_1+t_N)$.
For a more detailed derivation (worked out for $N=2$, but extending to arbitrary $N$) we refer to \cite{Efremidis2002},
only noting that the mass term in Eq.~(\ref{eq:NLS}) has the right form to ensure that near $k=0$ the dispersion curve of the effective continuum equation matches $\mu_0(k)$ of the DNLS model.
Now for $U<0$ and $\tilde \mu <0$, Eq.~(\ref{eq:NLS}) has exact bright soliton solutions of the form $\psi(x)=\beta \sech(\alpha x)$,
where $\beta=\sqrt{2 \tilde \mu/U}$, $\alpha=\sqrt{-\tilde \mu/(t_1+N^2 t_N)}$ \cite{Pethick2008}.
Their norm is thus given by
\begin{equation}
 \norm = \int_{-\infty}^\infty {\rm d}x|\psi(x)|^2 =\frac{4}{-U} \sqrt{-\tilde \mu (t_1+N^2 t_N) }.
\label{eq:NLSnorm}
\end{equation}
Equating the amplitude $\beta_{\rm cr}^2=2t_1\sin^2(\pi/N)/(-U)$ then yields
$\mu_{\rm cr} \approx -2(t_1+t_N) - t_1 \sin^2(\pi/N)$ as in Eq.~(\ref{eq:app1a}) of the text,
and inserting back into Eq.~(\ref{eq:NLSnorm}) gives Eq.~(\ref{eq:app1b}).
These results are compared to numerical data in Figs.~\ref{fig:var}(c,d).
In contrast to the variational ansatz, the continuum approximation becomes exact near the linear limit,
but increasingly fails for larger $|\mu|$, where it tends to underestimate the central peak and to overestimate the norm.

Comparing the two methods, the continuum approximation is more suited for highly delocalized breather solutions,
while the exponential variational ansatz tends to be more appropriate for slightly more localized modes.
Since the inverse localization length at the critical point scales like $a(\mu_{\rm cr}) \approx \frac{1}{2} \sin(\pi/N)/\sqrt{1+N^2 t_N/t_1}$ according to Eqs.~(\ref{eq:vara}) and (\ref{eq:app2a}),
the continuum approximation may thus be expected to be superior to the variational one for larger $N$ or larger $t_N$,
while the variational one is more suitable for small $N$ and $t_N$. Such a trend is indeed observable in Figs.~\ref{fig:N2}(f) and \ref{fig:N5}(d).

\section*{References}

\bibliographystyle{elsarticle-num} 
\bibliography{nonlinear}





%
%
%
\end{document}